\documentstyle[preprint,pre,aps,prl,texdraw,graphicx]{revtex}
\tightenlines
\begin{document}
%\draft 
\title{Effects of confinement on self-assembling systems}
\author { A. Ciach, V. Babin and M. Tasinkevych} 
\address{Institute of Physical Chemistry and College of Science, 
Polish Academy of Sciences \\
Kasprzaka 44/52, 01-224 Warsaw, Poland \\} 
\date{\today} 
\maketitle 
\begin{abstract}
Systems in which particles can self-assemble into mono- or bilayers can 
form 
variety of stable and metastable structures on a  nanometer length scale. 
For this reason confinement has a 
particularly strong effect on such systems. We discuss in some detail
 effects
 of confinement on lamellar and cubic phases with double-diamond structure.
Structural deformations in slit geometry are described for large and small
 unit cells of the structure (in units of the thickness of the monolayer)
 and for various strengths  of interactions with the confining surfaces.
We show how the structural changes of the confined fluid are reflected
 in the 
measurable solvation force between the confining walls.

\end{abstract}

%\pacs{PACS numbers: 68.15.+e, 68.55.-a, 68.55.Jk, 68.60.-p}

\section{Introduction}
Confinement plays a significant role when the size of the system
becomes comparable to a typical length characterizing the structure
of the confined fluid. Usually the typical length is related to an
average distance between particles, and the effects of confinement
manifest themselves is systems whose sizes are of order of several
molecular diameters. The characteristic lengths, such as the
correlation length, become large close to phase transitions,
and in such cases the effects of confinement are found in much
larger systems. Large typical lengths, often two orders of magnitude
larger than the molecular sizes, characterize the structure in
self-assembling systems. The typical length $\lambda$ corresponds
to the size of correlated domains (for example micelles or lamellae)
and is reflected in the form of correlation functions. On the length scale 
set by $\lambda$ the domains play a similar role as particles on the
 molecular length scale.
Because of the presence of the structure on the nanometer length
scale, the finite size effects are expected for system sizes
two or three orders of magnitude larger than in simple fluids,
even far from phase transitions.

Bulk phase diagrams in self-assembling systems are very rich
due to stability of liquid-crystalline phases with different
symmetry. Because phases with different structure, symmetry
and characteristic length are stable or metastable in the bulk,
the structure of the confined system may depend significantly
on the size and shape of the container.
The increase of the free energy of the stable or metastable phase,
which is associated with
 structural deformations
induced by the confinement, is much  larger 
if the bulk  structure 
is strongly incompatible with the shape and the size of the container.
 As a result it may happen
that the structure of the confined system is significantly
different (for example has a different symmetry) than the
structure of the bulk phase, because  some phase which is
 metastable in the bulk
may become stable in the confinement,  if its structure
fits better the geometry of the system.
 Such phenomenon resembles capillary 
condensation in simple fluids.

To summarize, we note that the structure on the nanoscale
in the self-assembling systems can play a role analogous to the
structure on the microscale in simple fluids, therefore
one can expect similarities between properties of the confined
complex fluids on the nanoscale and properties of confined
simple fluids on the microscale. On the other hand, various
stable and metastable phases with quite different structures
are present in the bulk self-assembling systems, in contrast
to the simple fluids. This difference between the complex and
the simple fluids may result in variety of phenomena in
confined self-assembling systems which in confined simple
fluids are absent.
\section{Self-assembling systems}
In the systems containing amphiphilic particles 
(lipids, surfactants, copolymers) one observes a spontaneous formation
of phases exhibiting a short- or a long-range order on a
length scale one or two orders of magnitude larger than the
size of the amphiphiles. Amphiphilic molecules
are typically elongated, and their two ends are of quite
different nature. One end is polar, whereas the other one
is non-polar and typically consists of a hydrocarbon chain.
When the amphiphiles are dissolved in a polar or a non-polar
solvent, then a self-assembly of amphiphiles into bilayers
takes place. This end of the amphiphile, which is repulsed by
the solvent, is hidden inside the bilayer, by which the
contacts between the polar and non-polar particles and
particle parts are avoided. Mixtures of polar and non-polar
liquids, such as water and oil, phase separate at room
temperatures. Amphiphiles added to such mixtures self-assemble
into monolayers, separating the oil- and water- rich regions.
The polar (non-polar) end of the amphiphiles forming the monolayer
is oriented towards the polar (non-polar) solvent
(Fig.\ref{fig1}).
Again, in ternary mixtures the unfavorable contacts between
the polar and non-polar particles or particle parts are avoided
when the self-assembly into monolayers occurs. Both the bilayers
and the monolayers can assume various shapes (Fig.\ref{fig1}).
The surface describing the center of the bilayers or
monolayers can be closed as in micelles or inverse micelles,
or not, as in lamellar and bicontinuous phases. The latter
phases are formed if there is no appreciable asymmetry
with respect to the shapes and interactions between the two
different parts of amphiphiles. In the lamellar phases the 
monolayers and bilayers fluctuate around their average positions,
described by flat, parallel surfaces. In the bicontinuous phases short-
or long-range order is present on the length scale set by the
size of the water-rich domains.
In the sponge-phase or in the microemulsion the actual
monolayers are described by surfaces characterized by very large
and negative Euler characteristics \cite{holyst:97:1}, 
but their average positions
are not localized.
The presence of the structure on the mesoscopic length scale
is reflected only in the form of the water-water density
correlation function \cite{strey:86:0,strey:87:0}. This correlation function 
exhibits a damped
oscillatory behavior with the period of oscillations $\lambda$
related to the size of the water-rich domains. Its form for
distances $\left.r\right/\lambda>1$ resembles
the density-density correlation function in simple fluids
for $\left.r\right/\sigma>1$, where $\sigma$ is a microscopic
length comparable to the size of molecules.
For $\left.r\right/\lambda<1$ the shape of the correlation
functions in microemulsion and sponge phases is different
than in simple fluids for $\left.r\right/\sigma<1$, and
reflects the fact that the domains are soft and compressible
rather then rigid. Hence on the mesoscopic length scale the
water-rich domains play a role analogous to particles on the
molecular length scale, except that the domains are soft
rather then rigid and under an external stress can change their shape
and volume.
There also exist bicontinuous structures exhibiting a long-range
order \cite{tabony:86:0,holyst:96:0}. In such phases the surfaces 
describing the average position
of the centers of bilayers or monolayers are periodic, that is a certain
unit cell is infinitely repeated in space. An example of such unit cell
is shown in Fig.\ref{fig2}. for the double-diamond (D) phase.

The self-assembly of amphiphiles into bilayers or monolayers, which
then assume different shapes on the mesoscopic length scale,
occurs in a large class of systems. The details of inter-particle
interactions in different systems exhibiting self-assembly can
be quite different. Despite these differences remarkable similarity
between properties of all such systems can be observed, provided
that one end of the solute particles (i.e. amphiphiles) attracts
polar and repulses non-polar particles, whereas the other end
does the opposite. This property of the interactions is necessary and 
sufficient
for self-assembly into bilayers or monolayers if the amphiphilic
 interactions are sufficiently strong, and the other details
of the interactions are irrelevant for the structure formation
on the mesoscopic length scale and for other  properties common
for all self-assembling systems, such as the very low surface tension
between some of the phases. The features common for the whole class
of the self-assembling systems are also shared by the members
of the class in which the inter-particle interactions are particularly
simple. The universal properties of the self-assembling systems
should be described by generic models in which the irrelevant
details of the interactions are just disregarded. Here we use the
generic model introduced by Ciach, H{\o}ye and Stell (CHS)\cite{ciach:88:0}
to study the universal effects of confinement.
\section{Model}
The CHS model \cite{ciach:88:0}  was introduced for description of balanced
ternary systems. In such systems symmetric amphiphiles form
monolayers with vanishing spontaneous curvature and the volume
fractions of oil and water are equal. The model can be easily
generalized for unbalanced systems. In particular binary
water-surfactant mixtures can be described by a generalized
version of the CHS model. Here we concentrate on the case with the
highest symmetry. We assume a symmetry with respect to a replacement
of the polar particles and particle parts by non-polar particles
and particle parts and vice versa. It means that if all the
amphiphiles change their orientations into the opposite ones
and simultaneously water particles replace the oil particles
and vice versa, the energy of the system is unchanged. In the
simplest case the model is defined on a simple cubic lattice
and close packing is assumed. The length unit is set by the lattice
 constant $a$, assumed to be equal to the length of the amphiphile. 
Each lattice cell can be occupied
by either a water, an oil or a surfactant particle. Different orientations
of the latter are treated as different components having the same
chemical potential. A typical configuration is shown in Fig.\ref{fig3}.
Non-vanishing interactions are determined by the polar-nonpolar
symmetry and by the interactions shown in Fig.\ref{fig3}.
We assume nearest-neighbor interactions. In the simplest version
of the model, except from the water-water (oil-oil) interaction $-b$
only the water-amphiphile (oil-amphiphile) interaction
$-c\Delta {\bf r\cdot{\hat u}}$ ($+c\Delta {\bf r\cdot{\hat u}}$) is
 assumed, with
${\bf {\hat u}}$ describing the orientation of the amphiphile located at
the distance $\Delta {\bf r}$ from the water (oil) particle.
In the extended model one assumes in addition the amphiphile-amphiphile 
interaction
$g[({\bf \hat{u}}\times({\bf r}^{\prime} - {\bf r}))\cdot
({\bf \hat{u}}^{\prime}\times({\bf r} - {\bf r}^{\prime}))]$, where
${\bf \hat{u}}$ (${\bf \hat{u}}^{\prime}$) is the orientation of the amphiphile
located at ${\bf r}$ (${\bf r}^{\prime}$). The above interaction supports
formation of flat monolayers, with amphiphiles parallel to each other
and perpendicular to the surface they occupy. If $g\neq 0$, the
stability region of the lamellar phase enlarges.

The CHS model can be further simplified if one requires that
${\bf \hat{u}}$ is reduced to $\pm{\bf\hat{e}}_i, i = 1,\dots,d$,
where ${\bf\hat{e}}_i$ are the unit lattice vectors. If only
one dimensional lamellar phase is stable for given thermodynamic
conditions, then ${\bf\hat{u}}$ can be projected onto the direction
${\bf\hat{n}}$ perpendicular to the lamellae (from water towards oil).
In this case one can distinguish two states of amphiphiles,
one with $({\bf\hat{u}}\cdot{\bf\hat{n}})>0$, the other one with
$({\bf \hat{u}}\cdot{\bf\hat{n}})<0$.

The model is relatively simple
and can be considered as a generic, semi-microscopic model
for the self-assembling systems. Having the property that one
end of the solute (amphiphilic) particles attracts water and
repulses oil and the other one does the opposite, the model
is a member of the class of the self-assembling amphiphilic
systems and can reliably describe those properties which are
shared by all the members of this class. Although quite simple,
the model can be solved exactly only in one-dimensional systems.
Therefore a mean-field (MF) approximation has been applied for
studying the bulk phase behavior and structure of various phases.
Fluctuations that are left out from the MF calculations
can change the quantitative results. Fluctuation-induced
first-order phase transitions \cite{buzano:97:0} can occur instead of continuous
transitions found in the MF approximation. Moreover, fluctuations
can destroy the order.
%The stability region of the locally ordered
%microemulsion can be enlarged at the cost of decreased regions
%of stability of the lamellar phase.
Monte Carlo simulations
performed for some other models of self-assembly \cite{gompper:93:0,holyst:97:1}
 show however
that the regions of stability of the ordered phases shrink
in the presence of fluctuations, but the ordered phases do
not disappear. Thus, the MF approximation gives qualitatively
correct phase diagrams for such complex systems.

The MF Hamiltonian for the CHS model has the form
\begin{eqnarray}
H^{\rm MF}=\sum\limits_{\bf r}
\left\{
\sum\limits_{i=1}^{2+2d}\phi_i({\bf r})
\left({\hat\rho_i}({\bf r}) - \frac{1}{2}\rho_i({\bf r})\right)
-\mu\left({\hat\rho_1}({\bf r}) + {\hat\rho_2}({\bf r})\right)
\right\},
\end{eqnarray}
\begin{eqnarray}
\phi_i\left({\bf r}\right)
=\sum\limits_{{\bf r}^\prime}\sum\limits_{j=1}^{2+2d}
u_{ij}\left({\bf r} - {\bf r}^\prime\right)
\rho_j\left({\bf r}^\prime\right),
\end{eqnarray}
where $\hat{\rho}_i({\bf r})$ is the microscopic density such
that $\hat{\rho}_i({\bf r}) = 1(0)$ if the site ${\bf r}$ is
(is not) occupied by the specie $i$, where $i$ refers to
water, oil and surfactant in different orientations.
$\rho_i({\bf r})$ is the equilibrium density of the state
$i$ at ${\bf r}$. The $u_{ij}({\bf r}-{\bf r}^{\prime})$ is
the described above interaction energy between the specie $i$
at ${\bf r}$ and the specie $j$ at ${\bf r}^{\prime}$.
Probability distribution for $\hat{\rho}_i({\bf r})$ is given
by the Boltzmann factor $P^{MF}\propto e^{-\beta H^{MF}}$, where
$\rho_i({\bf r})$ are to be found by solving self-consistent
set of equations
$\rho_i({\bf r}) = \left<\hat{\rho}_i({\bf r})\right>_{MF}$, where
$\left<\dots\right>_{MF}$ means averaging with $P^{MF}$. Self-consistent
solutions of the above equations are equivalent to finding local
minima of
\begin{eqnarray}
\Omega^{\rm MF}&=&
\frac{1}{2}\sum\limits_{{\bf r}, {\bf r}^\prime}\sum\limits_{i,j = 1}^{2+2d}
u_{ij}\left({\bf r} - {\bf r}^\prime\right)\rho_i({\bf r})\rho_j({\bf r}^\prime)
\nonumber \\
&+&\sum\limits_{\bf r}\left\{
kT\sum\limits_{i=1}^{2+2d}\rho_i({\bf r})\ln\rho_i({\bf r})
-\mu\left(\rho_1({\bf r}) + \rho_2({\bf r})\right)
\right\}.
\end{eqnarray}
The stable structure is identified with the global minimum
of $\Omega^{MF}$.

The CHS model predicts stability and coexistence of various
phases in the bulk. Already in its simplest, one dimensional version
oil-rich, water-rich, microemulsion and lamellar phases can be stable.
The lamellar phases range from phases with short periods
(high surfactant concentration) to highly swollen phases. When
6 orientations of amphiphiles, corresponding to
$\pm\hat{\bf e}_i, i=1,2,3$ are allowed, then a cubic
bicontinuous phase D with double-diamond symmetry is stable
in addition to the above mentioned phases. For continuous $\hat{\bf u}$
also cubic G (gyroid), P (simple cubic) and close-packed micelles and 
reverse micelles occur. Because of the rich phase behavior and
relative simplicity of interactions, the model can be applied
for studying the effects of confinement on the microemulsion,
lamellar and cubic phases. The results obtained within the MF
approximation for the lamellar and the D phases are described
below.
\section{Slit geometry}
We will focus on two parallel walls between which the system
under consideration is confined. We assume that the distance $L$
between the walls is small i.e. $L^2\ll A$, where $A$ is the area
of each wall (Fig.\ref{fig4}). The grand thermodynamic potential
of the confined system has the form
\begin{eqnarray}
\Omega=\omega_b A L + \Omega_{ex},
\label{sigma}
\end{eqnarray}
where $\omega_b$ is the bulk grand-thermodynamic potential
density, and where $\Omega_{ex}$ is the excess potential
due to the presence of the confining walls. The confinement leads to
 additional terms in $d\Omega$, which for the slit
geometry is given by
\begin{eqnarray}
d\Omega = - SdT - Nd\mu - pdV +2\sigma dA - f AdL,
\end{eqnarray}
where $\sigma$ is the wall-fluid surface tension, and
$fA$ is the force which has to be applied externally to keep
 the walls at the distance $L$,
\begin{eqnarray}
f=-\frac{1}{A}\left(\frac{\partial\Omega}{\partial L}\right)_{\mu, T, A}-
p=-\frac{1}{A}\left(\frac{\partial\Omega_{exc}}{\partial L}\right)=
f_{ww}(L)+f_s(L,\mu,T).
\end{eqnarray}
The $f$ consists of the direct force between the walls, $f_{ww}$, present
for $L$ comparable to a range of intermolecular interactions, 
and of the solvation force, $f_s$,
induced by the confined fluid. In simple fluids the solvation force
reflects packing effects of particles \cite{israel:81:0}. When in the confined
liquid the average distance between particles is larger (smaller)
than in the bulk, then attraction (repulsion) between the
confining walls results, since the confined fluid tends to
assume the equilibrium bulk structure. When the wall separation
increases and the average distance between particles starts
to exceed 1.5 times the bulk equilibrium distance, a new
layer of particles is introduced into the slit and the attractive
solvation force abruptly changes into the repulsive one.
Due to the analogy between the particles on the microscale
and the water-rich domains on the mesoscale, similar packing
effects of domains can be expected on the length scale set by
the size of the domains. As the size of the domains is
$\propto 10-100$ nm, the packing effects of domains should lead
to oscillating solvation force (repulsive for compressed domains
and attractive for expanded domains) for quite large separations
between the confining walls, even up to micrometers.
On the other hand,  one cannot expect complete analogy
between the confined self-assembling system on the mesoscale and the 
confined simple
fluid on the microscale. An important difference
between the simple fluid on the microscale and the self-assembling
system on the mesoscale is the presence of variety of stable and
 metastable phases with different symmetries and sizes of the unit cell
 in the latter. As we show later, a symmetry of the ordered phases 
plays a very important role for the form of the solvation force.
 Also, various deformed structures can occur, since
 the domains are flexible and can change the shape and volume, 
unlike the particles, which are rigid and of a fixed size.

\section{Confined lamellar phases}
\subsection{strongly hydrophilic walls}

Walls which are strongly hydrophilic attract polar particles or particle
 parts. Therefore either a water-rich or an amphiphilic layer is adsorbed near
 each wall, by which the orientation of the subsequent layers is fixed. 
One can expect that the average densities of all the components are 
constant in the planes
 parallel to the confining walls and the density profiles $\rho_i(z)$ 
depend only on a distance $z$ of one arbitrarily chosen wall, say 
the left one. [Later we show that this assumption is wrong in the case of
 lamellar phases with very short periods.] Because the orientation of the 
lamellae is fixed, one can 
consider the simplest version of the CHS model, with only 4 states, namely
 water, oil and surfactant with the head or the tail oriented towards the 
left wall. The interactions between the $i$-th  component and the wall are 
assumed to be the same as between the $i$-th component and water. In fact we 
consider water-covered walls. 
\subsubsection{swollen lamellar phases}

The swollen lamellar phases confined between strongly hydrophilic walls 
have been studied experimentally by the surface force apparatus (SFA) 
measurements 
\cite{kekicheff:89:0,kekicheff:90:0,kekicheff:91:0,kekicheff:97:0,abillon:90:0,petrov:94:0}. The results of the experiments show that for the
 distance between the walls $L\ge 5\lambda$, where $\lambda$ is the period 
of the bulk lamellar phase, the film responds elastically to the applied 
stress. For $N$ periods of the lamellar phase the confined fluid
behaves in the same way as a series of $N$ joint identical 
springs with low elastic modulus. 
Such behavior was observed for different substances, 
and should be thus reproduced in any generic model of self-assembling 
systems. 

We begin our study of the effects of confinement on the 
self-assembling systems by considering the swollen lamellar phases confined
between parallel hydrophilic walls. This will serve as a test of the 
applicability of the CHS model for confined water-oil-surfactant mixtures. 
 We choose model parameters such that
the period of the bulk phase in the units of the thickness of the monolayer,
 $a$, is $\lambda\approx 10$ \cite{tasinkevych:99:2},
 since it is in the same range as 
 the periods in the experimentally studied systems
 (in the same physical units) \cite{kekicheff:97:0}. 
 The excess thermodynamic potential per unit
 area, $\Omega_{ex}/A$, and the solvation 
force, $f$, obtained  for $\lambda=13$ in Ref.\cite{tasinkevych:99:2} 
are shown in Fig.\ref{fig5}. For $L=L_N=N\lambda +\lambda/2+1$ (in $a$ units) 
the $\Omega_{ex}$ 
assumes minima, which correspond to $N$ lamellar layers confined between 
the walls, and the period in the confined system is the same as in the bulk
 equilibrium.
 For $L\ne L_N$ the periods of the confined 
and the bulk phase 
are different at the same thermodynamic conditions. We should stress that in
 the case of the swollen lamellar phases the water (and oil) rich layers are
 sufficiently thick to be compressible in a way similar to the bulk fluid. 
This fact 
corresponds to metastability of lamellar phases with different periods 
(different thicknesses of the water- and oil-rich layers) in the bulk.
 The metastable phases   become stable
 under external stress, i.e.  between the parallel walls a distance $L\ne L_N$
apart. When the slit is expanded, $L_N\to L_N+ \Delta L$, then a new lamellar
 layer is introduced into the slit for $\Delta L>\lambda/2$, and the 
attractive solvation force abruptly changes into the repulsive one 
(see Fig.\ref{fig5}.b).
  
For  large separations between surfaces $(N>4)$ the second derivative of 
$\Omega_{ex}$ with respect to $L$, calculated at
$L=L_{N}$, $B=\Omega_{ex}^{\prime \prime}(L_{N})$,
 is well approximated by a 
straight line $B=\bar B/L_{N}$ as a function of $1/N$ 
(see Fig.\ref{fig6}), where $\bar B$ is a modulus of compressibility. Hence the
response of the system to compression or decompression is elastic, and
analogous to the behavior of a series of identical joined springs.
$\bar B$ as a function of
 $P=\lambda/2$  for a three-dimensional system is shown in 
Fig.\ref{fig7} (open circles) together with the 
phenomenological curves \cite{helfrich} for  two values of a membrane rigidity,
$\kappa$,
 between which 
the results are located \cite{tasinkevych:99:2}. The phenomenological 
curves are obtained for a lamellar phase modeled  as a stack of elastic,
 undulating membranes  with the
 bending elastic modulus $\kappa$. The membranes represent 
 surfactant monolayers. The discrepancy between 
the results of the CHS and the membrane models results probably from 
deformations, such as passages or droplets, neglected in the latter case.
The agreement with experiments \cite{kekicheff:97:0}, on the other hand,
 is very good. 
We obtain semiquantitative agreement between the model and the 
experimental results once the periods of the bulk phase in units of 
the thickness of the monolayer are the same in the two cases 
\cite{tasinkevych:99:2}. 
Note that
the model parameters are chosen such that the bulk structures in the model
 and experiment agree, and then no parameters are fitted in the case of 
the confinement.

For the  surface separations for which 
the number of adsorbed layers is $N\leq 3$, 
the stretch strain of layers  releases by formation of the uniform 
water-rich film for $N=1,3$ (see Fig.\ref{fig8}.a),  and oil-rich film for $N=2$ (see
 Fig.\ref{fig8}.b) in 
the middle of the slit. The formation of the uniform 
films inside the slit is reflected  in the saturated-like behavior
of $\Omega_{ex}$ and considerably low $ f$ for the  corresponding 
 surface separations. For such separations the  central, uniform layer 
grows and the near-surface lamellar structure remains unchanged
 when the slit is expanded. 
The growth of the uniform water- or oil-rich central layer is not
 accompanied by the elastic response. Note that for $L<5\lambda$ removing
 one layer, $L\to L-\lambda$, results in a large deformation, that is 
$\lambda\to \lambda +\Delta\lambda$ with $ \Delta\lambda/\lambda >20\%$. 
On the other hand, 
the swollen lamellar phases are stable close to the coexistence
 with the uniform water-/oil-rich phases
 (water-oil symmetry was assumed). Close to the coexistence the 
bulk density of the grand
 thermodynamic potential in the stable lamellar phase and the metastable
 oil- or water-rich phases  are only slightly different.
 Moreover, the surface tension between
 the uniform and the swollen lamellar phases is very low. Thus, when 
the near-surface lamellar
 films have the same structure as the bulk phase, the cost
 of nucleation   of a uniform layer in
the center of the slit is lower, 
than the elastic-energy cost associated
with  large deformation, $ \Delta\lambda/\lambda >20\%$, of the 
lamellar structure throughout  the whole slit. 
\subsubsection{Short-period lamellar phases}

The short-period lamellar phases are stable for rather high surfactant 
concentration. The water- and oil-rich layers are thin in this case, 
and do not behave as the bulk fluid layers, in contrast to the layers in the swollen 
lamellar phases. When the layers of water and oil are incompressible,
  the phases
 other than the lamellar phase with one specified period are unstable in the
 bulk. Actually, all the phases except for the lamellar phase with $\lambda=4$ 
are unstable in a certain region of the phase space in the CHS model
\cite{tasinkevych:99:2}. 
The lack of metastability of the other phases has a strong effect on the 
structure of the confined lamellar phase, especially when the size of the 
system and the period of the only stable phase are incompatible. In the absence
of metastable phases with different periods we cannot expect that the period
of the confined phase will increase or decrease under expansion or 
compression of the slit. Also, the nucleation of a layer of 
a uniform phase in the center of the slit, found in the case of the swollen
 lamellar phases, cannot take place when the uniform phases are not
 metastable in the bulk. Hence, some other kinds of deformations can be
 expected. The   short-period lamellar phases resemble smectic liquid 
crystals, which show Helfrich \cite{helfrich:70:0,hurault:73:0,delaye:73:0,clark:73:0} 
undulation instability under external
 stress. We should thus take into account a possibility of a similar behavior 
in oil-water-surfactant mixtures. The undulating structures are no longer 
one-dimensional, and we shall
thus assume, as in Ref.\cite{tasinkevych:00:1} that the 
densities have at least 
a two-dimensional structure and depend on
two space variables, $(x,z)$ (in directions given by $\hat {\bf e}_i$
 with $i=1,3$), where $z$ is a distance from one wall and $x$ is 
a coordinate in the direction parallel to the walls. We assume that 
in the other  direction parallel to the walls the 
densities are constant. When the orientation of the lamellae is no longer
 fixed, then the assumption of only two relevant orientations of the amhiphiles
is no longer justified. We thus assume that the orientations of the amphiphiles
are restricted to $\pm \hat {\bf e}_i$ with $i=1,3$, when the densities change
 in the two corresponding directions.

We choose the model parameters and the
 thermodynamical state such that the lamellar phase with the period 
$\lambda=4$ is stable and no metastable phases are present
\cite{tasinkevych:00:1}.  By numerical minimization of $\Omega_{ex}$ we
 found in Ref.\cite{tasinkevych:00:1} that the 
structures corresponding to the global minimum of $\Omega$ for $L\ne L_N$ 
show the following pattern. For $L=L_N+1$ the central water- or oil-rich 
layer is twice as thick
 as the corresponding boundary layers. For $L=L_N+2$ the structure shown in 
Fig.\ref{fig9}.a occurs. This structure 
strongly resembles the Helfrich undulation instability in smectic 
liquid crystals. For very thin layers of oil and water between the monolayers 
of amhiphiles the lamellar phase should indeed be very similar to a smectic
 liquid crystal. For  
$L=L_N+3$ we observe a formation of lamellar layers  near each
 wall, whereas in the center of the slit the lamellae are oriented
 perpendicularly to the confining walls (Fig.\ref{fig9}.b). For the perpendicular 
orientation there is no constrain
 on the period of the lamellar phase within the central lamellar layer.
 The structures shown in Fig.\ref{fig9}. consist of
domains of the lamellar phase having  the bulk structure, and of layers of
 deformations separating them. The thickness of the layers of
 deformations is comparable to 
the lattice constant, which in this case is the characteristic structural
 length. 
\subsection{Neutral and weakly hydrophilic walls}

When the confining walls are strongly hydrophilic, the  lamellae are
 parallel to them, since
the wall-fluid surface tension in a semi-infinite system is lower for this 
orientation than for any other one. For other orientations of the lamellae
 either the unfavorable 
contacts between the polar and nonpolar particles and/or particle parts, or
 deformations of the structure cannot be avoided. For a perfectly neutral
 wall, however, neither component nor orientation of amphiphiles is favored 
near the wall and one cannot a priori assume that the lamellae should be 
parallel to the walls. In fact in the previous section we have observed that 
a perpendicular orientation of lamellae stabilizes in the center of the slit, 
by which the deformations of the lamellar structure  within the central
layer are avoided. 
The lamellar layers formed near each wall can be considered as a kind of
 'external walls' for the central layer of perpendicularly oriented lamellar
 phase. The water concentration at the surface which is in contact with the 
central layer is lower than at the
surface covered by pure water, and this surface can be considered as a weakly 
hydrophilic one. Lamellae perpendicular to very
weakly hydrophilic walls were also observed in  Monte Carlo simulations of a
Landau model of microemulsions \cite{holyst:98:0}. Based on the above observations
we can expect that the
 perpendicular orientation of the lamellae is  more favorable than the
 parallel one for very weakly hydrophilic or perfectly neutral walls.
 We have found \cite{tasinkevych:00:0} that indeed, the perpendicular 
orientation stabilizes between neutral walls, but only for small periods of
 the lamellar phase, which correspond to the surfactant volume fraction
 $\rho_s\ge 1/3$. For swollen lamellar phases we always find stability of
 lamellar phases  parallel to the walls in the CHS model
\cite{tasinkevych:00:0}. 

In order to study the case of weakly hydrophilic walls, we have 
to define the interactions between the surface which is considered as 
a  weakly hydrophilic one and all the components of the mixture.
  We assume that the
interactions between each specie in the mixture and the wall are uniformly
decreased compared to the interactions between this specie and the water
particle. To obtain the interactions between each component and the 
weakly hydrophilic wall 
we multiply  the bulk interactions between each component and the
water particle  by the same factor $0<h_s<1$.

When the walls are  weakly hydrophilic (or hydrophobic) then in a 
semi-infinite system the parallel orientation of
the lamellar phase is preferable, because the unfavorable 
 contacts between the polar
and the nonpolar particles or particle parts are avoided,
 in contrast to the perpendicular orientation of the lamellae.
 On the other hand, when
the width of the slit and the period of the lamellar phase do not match,
 the lamellar structure parallel to the walls 
is deformed. The deformation of the structure 
 leads to the elastic contribution to the free energy, as in
the case of the strongly hydrophilic walls. 
For the parallel orientation $\Omega_{ex}$ contains the
 surface tension contribution $\sigma_{\parallel} A$ and
the elastic energy contribution $\bar B(L-L_N)^2/2L_N$. The latter vanishes 
  in the case of no stress ($L=L_N$) 
and assumes a maximum for the most deformed
structure ($L-L_N=\lambda/2$). When
 the lamellar phase is oriented perpendicularly to the
confining walls  there is
no constrain on the period of the structure, which can be the 
same as in the bulk  and
there is no elastic contribution to $\Omega_{ex}$, i.e.
 $\Omega_{ex}\approx \sigma_{\perp}A$.     
The surface tension in this case
is larger than in the case of the parallel orientation. In the absence of
the deformations of the structure,  i.e. for $L=L_N$, 
the parallel orientation should be thus
stable in the slit. However, when the elastic modulus $\bar B$ is 
sufficiently large, the
elastic energy of deformations may be larger than the difference between
the surface tensions $\sigma_{\perp}- \sigma_{\parallel}$ 
for wall separations $L\ne L_N$, corresponding to a strong
stress. For such wall separations 
 a switch to the perpendicular orientation may occur.

We have observed this switch in our model \cite{tasinkevych:00:1,tasinkevych:00:0}
 for the surfactant volume
fraction $\rho_s=1/3$ and for very weakly hydrophilic walls, for example for
 $h_s=0.015$ (Fig.\ref{fig10}). 
Lamellar phase with the surfactant volume fraction  $\rho_s=1/3$ 
is stable at room
temperatures for example in water, decane and $C_{10} E_5$ \cite{kahlweit:86:0}. 

The switch should lead to an abrupt change of electrical conductivity,
reflectivity, elastic properties (large solvation force for the parallel
and almost vanishing for the perpendicular orientation). Thus by slight 
changes of the control parameter such as the width of the slit we can 
induce abrupt, qualitative  changes of various 
physical properties of the system.

\section{Confined cubic bicontinuous structure with double-diamond symmetry}

The cubic phase with the double-diamond structure, D, is stable in the CHS 
model with 6 allowed orientations of amphiphiles, $\hat u=\pm \hat{\bf e}_i$, 
$i=1,2,3$. For sufficiently strong amphiphilic interactions $c/b$ the D
 phase is stable in an extended region of the $(\mu_s,T)$ phase diagram
 ($\mu_s$ is the surfactant chemical-potential and $T$ is the temperature). 
The size of the unit cell, $\lambda$, can be large close to the coexistence
 with the oil- and the water-rich phases. The $\Omega_{ex}$ was obtained 
 in \cite{babin:00} for a 
system confined between water-covered walls and 
for the parameters corresponding to a stability of the bulk phase 
with the size of the unit cell $\lambda=8$ (Fig.\ref{fig11}).
Note that for sufficiently large $L$ the period of $\Omega_{ex}$ is 
equal to  $\lambda/4$, whereas in the case of the lamellar phase the period of 
$\Omega_{ex}$ is equal to the period of the bulk lamellar structure. 
The relation between the characteristic length of the structure, 
which for the ordered phases can be identified with the size of the unit 
cell $\lambda$, and the response of the confined system to the external
 stress is 
thus essentially different for the two phases. This difference follows 
from different symmetries of the phases in question. Note that in the case
of the lamellar phase the whole period of the structure has to be introduced 
into the slit, if the unfavorable contacts between the polar and nonpolar 
particles and/or particle parts are to  be avoided. It is not the case for
 the D structure, because it has a different symmetry. To see the role of the 
symmetry of the D phase, let us examine Fig.\ref{fig12}, where the central parts of 
the water- and the oil-rich channels are shown schematically for the unit cell.
 The D structure
consists of the water- and the oil-rich channels and the junctions between
 them. In each junction 4 water-rich or 4 oil-rich channels ('legs')
 are connected
 with each other and form a tetrahedron. There are 4 layers of junctions 
in the unit cell
(Fig.\ref{fig12}). In each layer of junctions there are 2 water and 2 oil junctions.
 The structure in each layer of junctions can be obtained from the 
structure of 
the previous layer by a symmetry operation (a rotation and a 
translation). The interaction energy between
the infinite hydrophilic surface and  the infinite periodic structure
 is thus the same regardless of which layer  of junctions forming the
 unit cell is exposed to the surface
(Fig.\ref{fig12}). Therefore under the expansion of the slit a single
layer of junctions is introduced into the system, by which large structural 
deformations are avoided. Since the thickness of the single layer of 
junctions is $\lambda/4$, also the period of $\Omega_{ex}$ is equal to
 $\lambda/4$, as we have found by numerical minimization of  $\Omega_{ex}$ 
in the CHS model.
Because the period of  $\Omega_{ex}$ is small compared to  $\lambda$, we 
have to consider large unit cells to verify whether the response to the 
compression or expansion is elastic and how large (or small) 
is the elastic modulus
compared to the lamellar phase with the same period. This work is in progress
\cite{babin:00}.
For other cubic phases we expect similar relation between the period of 
 $\Omega_{ex}$ (and also the measurable  solvation force) and the symmetry
 of the unit cell of the bulk structure.

Interesting structural deformations occur for narrow slits, $L< 2\lambda$.
First, for $L<\lambda$ a capillary condensation of water, which is stable in 
the bulk for slightly smaller $\mu_s$, occurs. For $\lambda<L<\frac{5}{4}\lambda$ a single layer of 
parallel
oil-rich channels is formed. In Fig.\ref{fig13} the oil-water interface is shown for
$L=\frac{9}{8}\lambda$. Formation of a layer of cylinders ('wheels') can be important for
lubrication of the two surfaces put into motion in direction perpendicular 
to the axes of the cylinders.
 Further expansion of the slit leads to
 a formation of  two layers of channels near the two walls.
  In each layer the channels are parallel to each other and perpendicular to 
the channels  in the other layer.  The oil-rich channels perpendicular to each
other are connected in 
a single layer of junctions, which is formed in the center 
of the slit (Fig.\ref{fig14}). The resulting  quasi two-dimensional bicontinuous structure 
looks like a nanoscopic sieve. 
 As in the bulk D phase, there are 4 'legs' merging 
in each junction. Unlike in the bulk, however, the channels are nearly 
parallel to the external walls, whereas in the bulk they form a tetrahedron. 
 
For larger wall separations a second layer of junctions of the oil-rich 
channels
is formed, and then further expansion leads to a formation of a third,
 fourth etc. layer of junctions. The channels located close to the confining 
walls remain parallel 
to them for arbitrary $L$. For an odd (even) number of layers of junctions 
in the confined 
system the channels near one external surface are perpendicular (parallel) to 
the channels near the other external surface. We expect such a behavior for
D phases with large periods, i.e. for relatively small $\mu_s$. 

\section{Summary}

We reviewed the results obtained in the CHS model for the lamellar
 and the D phases 
confined between two parallel walls 
\cite{tasinkevych:99:2,tasinkevych:00:1,tasinkevych:00:0,babin:00}. The results
 show that there is indeed a similarity between the confined
complex fluids on the mesoscopic length scale
 and the confined simple fluids on the molecular length scale. Namely, the
solvation force oscillates in each case with the periodicity related to a
typical structural length. While in the simple fluids the typical length is
related to the average distance between particles, in the case of the complex
 fluids the typical structural length  depends  on the symmetry and the
 anisotropy of the ordered phase. In the case of the lamellar phase the 
characteristic structural length is  equal   to the period of the lamellar 
phase, 
i.e. to the size of the unit cell $\lambda$. In the case of the D phase
  the structural unit corresponds to the single
 layer of junctions of the water- and oil-rich channels, and its thickness is
equal to a quarter of the size of the unit cell,  $\lambda/4$.

There are other important differences between the confined simple fluids and 
the confined complex fluids with different symmetries. 
Let us first describe  the
 differences related to 
the  anisotropy of the ordered phases. The anisotropy plays 
an important role for the slit geometry. The elastic properties of the film of
an anisotropic sample depend on the orientation of the anisotropy axis with 
respect to the confining walls. Here we have described a switch of the 
lamellae orientation induced by compression or expansion of the lamellar 
film confined between very weakly hydrophilic walls. This phenomenon is 
associated with the 
anisotropy of the lamellar phase. 

Another important property of the confined 
self-assembling systems is the close relation between the structures which 
can be stabilized between the confining walls and metastability of various 
phases in the bulk. We have described this relation in some detail in the 
case of the lamellar phases. The structure of the confined system is 
determined by
a single phase metastable in the bulk if its structure fits the size of
 the slit and is not much different than the structure of the bulk phase 
(i.e. $\omega_b$ has almost the same value for the two phases). In the case 
of stabilization in the slit of metastable lamellar phases with different 
periods for different $L$, 
 we observe an elastic response to the 
applied stress.

 For narrow slits it turns out that a
nucleation of a single layer of a uniform phase between two near-surface
 lamellar layers is more preferable
 than stabilization of a highly deformed lamellar structure in the whole slit.
The lamellar phase with a significantly different period 
 has a significantly 
higher  $\omega_b$ than the equilibrium bulk phase. The uniform phase 
 is also metastable 
in the bulk, and the surface tension between it and the lamellar phase is 
low. For more complex shapes 
of the confining walls one should thus take into account a possibility of 
stabilization of two or more metastable phases separated by interfaces, if
 there are many metastable phases in the bulk and the surface tensions
 between some of them are low. 

 If there are 
no metastable phases in the bulk, essentially different behavior is found in 
confinement. Domains of the bulk structure are formed, and the shapes of the 
boundaries and the orientations of the domains are adjusted to the size and 
the shape of the container. A confined system with dislocation-like
 deformations or with domains of a uniform structure does not respond 
elastically to the applied stress.

For the D phase we find stabilization of elementary structural units in the
 narrow slits. First a layer of channels is formed, then a single layer of
 junctions. For other cubic bicontinuous phases we may expect similar
 stabilization of elementary units.

The properties of the confined self-assembling systems may be important
 for various applications. Templates for various ordered structures are formed
spontaneously and one can control the structures by controlling the 
thermodynamic conditions. Some of the structures, such as the layer 
of cylinders or the sieve-like structure, as well as transitions between 
them, such as the switch of the lamellae orientation, may be important 
for modern technologies.

{\bf ACKNOWLEDGMENTS}

This work was partially supported by the Polish State Committee for Scientific
Research (grant 3 T09A 073 16).
\pagebreak

%%%%%%%%%%%%%%% Fig.1 %%%%%%%%%%%%%%%%%
\begin{figure}
\caption{Schematic representation of the self-assembly in the ternary 
oil-water-surfactant mixtures.}
\label{fig1}
\end{figure}
%%%%%%%%%%%%%%% Fig.2 %%%%%%%%%%%%%%%%%
\begin{figure}
\caption{Unit cell of the cubic D phase (the oil-water interface, separating
the oil- and water-rich domains is shown).}
\label{fig2}
\end{figure}
%%%%%%%%%%%%%%% Fig.3 %%%%%%%%%%%%%%%%%
\begin{figure}
\caption{CHS lattice model. Top: a typical configuration. Bottom: non-vanishing interactions.}
\label{fig3}
\end{figure}
%%%%%%%%%%%%%%% Fig.4 %%%%%%%%%%%%%%%%%
\begin{figure}
\caption{Slit geometry.}
\label{fig4}
\end{figure}
%%%%%%%%%%%%%%% Fig.5 %%%%%%%%%%%%%%%%%
\begin{figure}
\caption{a: the excess thermodynamic potential $\Omega_{ex}$ per unit surface area
(in units of ${ b}/a^2$), defined 
in
 (\ref{sigma}), as a function 
of the wall separation measured in units of the lattice constant $a$.
b: the solvation force ${ f}$ (in units of ${ b}/a^3$) as a function 
of the wall separation. The thermodynamic variables
 $\tau=kT/b$, $\mu/b$ and the material constant ${ c/b}$ of the one-dimensional system
correspond to  stability of the swollen lamellar phase with $\lambda=13a$ 
 $(\tau=0.84, \mu/b=0.774, { c/b}=1)$; the distance from
the first-order transition between the water-rich and the
lamellar phases is $|\Delta\mu/b|=0.003$. Walls are covered by water.
 Dashed lines are to guide the eye.
}
\label{fig5}
\end{figure}

%%%%%%%%%%%%%%% Fig.6 %%%%%%%%%%%%%%%%%
\begin{figure}
\caption{The thermodynamic variables $\tau$, $\mu/b$ 
and the material constant ${ c/b}$ are the same as in  Fig.\ref{fig5}.  The subsequent minima of $\Omega_{ex}$ are fitted by quadratic curves
 $ B(L-L_N)^2$, where $L_N$ is the equilibrium separation for
$N$ adsorbed layers and $B$ is the coefficient in the 
fitting curve  to the $N$-th minimum.   
$B$ (in units of ${ b}/a^4$) is shown  as a function of the inverse number of adsorbed 
layers $1/N$. 
Dashed line is a linear fit.
}
\label{fig6}
\end{figure}
%%%%%%%%%%%%%%% Fig.7 %%%%%%%%%%%%%%%%%
\begin{figure}
\caption{Compressibility
modulus $\bar B$ (in units of $kT/a^3$), 
of the swollen lamellar phases as a function of the 
average distance between the surfactant monolayers, $P=\lambda/2$ 
(in units of $a$) in the three-dimensional system.  The
behavior expected from the phenomenological prediction of Helfrich 
%\cite{helfrich}
 is shown
for two  values of $\kappa$: $0.5kT$ (solid line),
$0.6kT$ (dashed line).}
\label{fig7}
\end{figure}
%%%%%%%%%%%%%%% Fig.8 %%%%%%%%%%%%%%%%%
\begin{figure}
\caption{One-dimensional system in the case  of intermediate wall separations,
$18<L<54$. Length is measured in units of the lattice constant $a$. 
The thermodynamic variables $\tau$, $\mu/b$ 
and the  material constant ${ c/b}$ are the same as in  Fig.\ref{fig5}, period of the bulk lamellar phase is $\lambda=13$. Walls are covered 
by water.
a: the density distribution of  water between the walls for the separation 
$L=29$.
b: the density distribution of  oil  between the walls for the separation 
$L=39$.
}
\label{fig8}
\end{figure}
%%%%%%%%%%%%%%% Fig.9 %%%%%%%%%%%%%%%%%
\begin{figure}
\caption{The oil-water interfaces, $\rho_1(x,z)-\rho_2(x,z)=0$, are shown
 in the $(x,z)$ plane,
where $\rho_1,\rho_2$ are the densities of water and oil respectively.
The white (dark) regions represent the water-rich (oil-rich) domains.
 The values
of the thermodynamic variables and the coupling constants are
$\tau=kT/b=2.5 $, $\mu/{ b}=1$, ${ c/b}=2.5$, ${ g/b}=1$ and $h_s=1$ and
correspond to a stability of the lamellar phase with the period $\lambda=4$
(in $a$ units).
a: the distance between the walls is equal to $25$ (in $a$ units).
b: the distance between the walls is equal to $26$ (in $a$ units).}
\label{fig9}
\end{figure}
%%%%%%%%%%%%%%% Fig.10 %%%%%%%%%%%%%%%%%
\begin{figure}
\caption{Excess grand thermodynamic potential for the parallel, $\Omega^{ex}_{\parallel}(L)$ (white circles),  and the perpendicular, 
 $\Omega^{ex}_{\perp}(L)$ (triangles down), orientations
as a function of the wall separation (in units of $a$) for $ c/b\rm =2.4$, $kT/b=\rm 2.8$, $\mu/ b=\rm 3$ and $ g/b\rm =0.15$.
 (a) $h_s=0$ (neutral walls);
(b) $h_s=0.015$ (weakly hydrophilic walls); (c) $h_s=1$ (strongly hydrophilic walls). }
\label{fig10}
\end{figure}
%%%%%%%%%%%%%%% Fig.11 %%%%%%%%%%%%%%%%%
\begin{figure}
\caption{$\Omega^{ex}$ for the confined D-phase as a function of the distance $L$ 
between confining walls (measured in $a$ units). Dark region corresponds to the water condensation in the slit. The parameters $kT/b=2.3, \mu/b=3.5, c/b=2.5,$ 
correspond to stability
of the bulk D-phase with the size of the unit cell $\lambda=8$ (in $a$ units).}
\label{fig11}
\end{figure}
%%%%%%%%%%%%%%% Fig.12 %%%%%%%%%%%%%%%%%
\begin{figure}
\caption{Schematic representation of the D phase. White (dark) cylinders represent 
water (oil) channels and the spheres represent the junctions. Cross-sections 
of the $4$ layers of junctions are also shown  with 
junctions indicated as dark and white circles.}
\label{fig12}
\end{figure}
%%%%%%%%%%%%%%% Fig.13 %%%%%%%%%%%%%%%%%
\begin{figure}
\caption{Oil-water interface of the structure formed in the slit for $L=\lambda+\lambda/8$.}
\label{fig13}
\end{figure}
%%%%%%%%%%%%%%% Fig.14 %%%%%%%%%%%%%%%%%
\begin{figure}
\caption{Oil-water interface of the structure formed in the slit. $L=\lambda+3\lambda/8$.}
\label{fig14}
\end{figure}

%
% Actual figures
%

\LARGE
\pagebreak
\begin{center}
Fig.1

\includegraphics[scale=0.6]{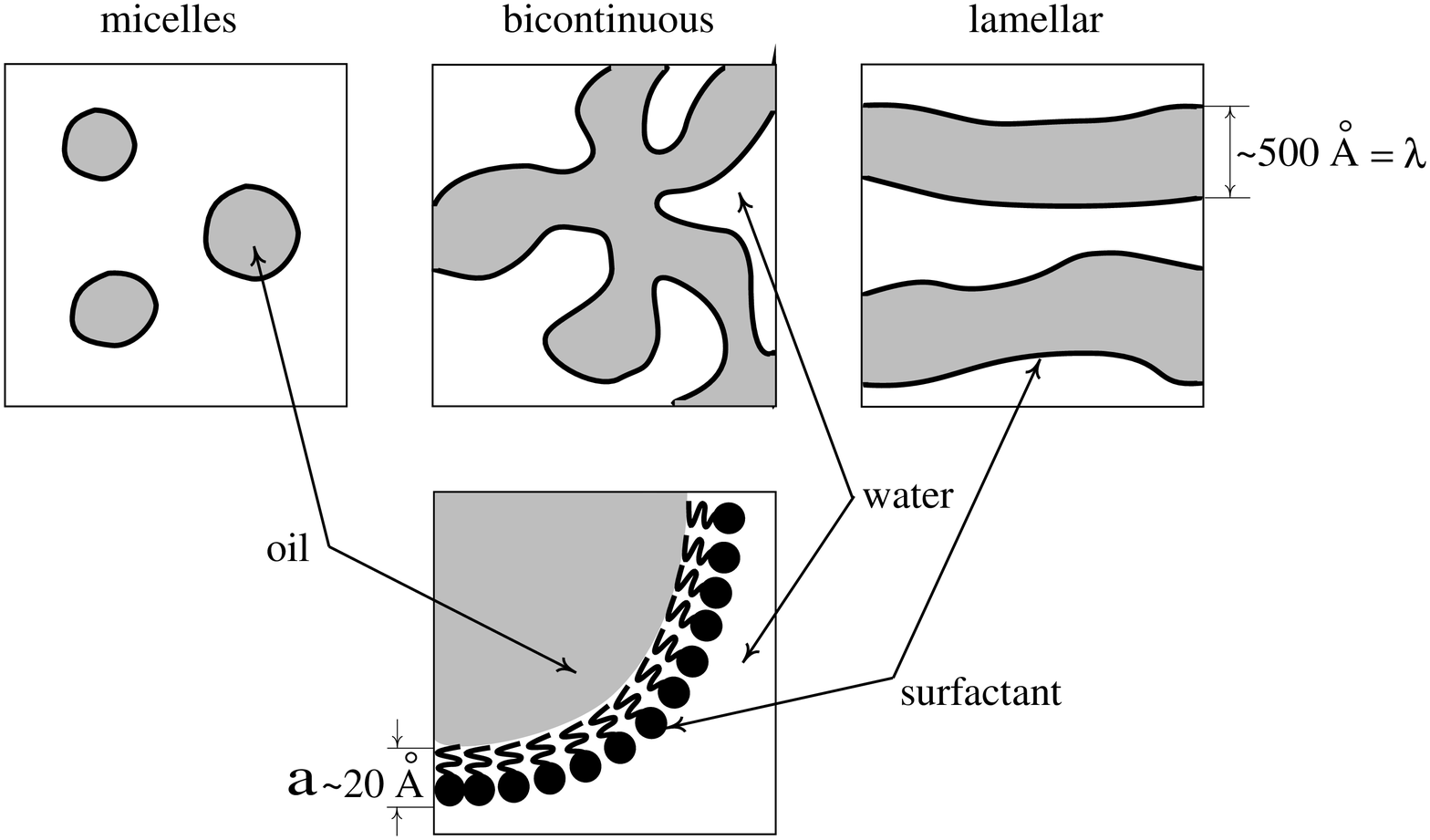}
\end{center}

\pagebreak
\begin{center}
Fig.2

\includegraphics[scale=1.0]{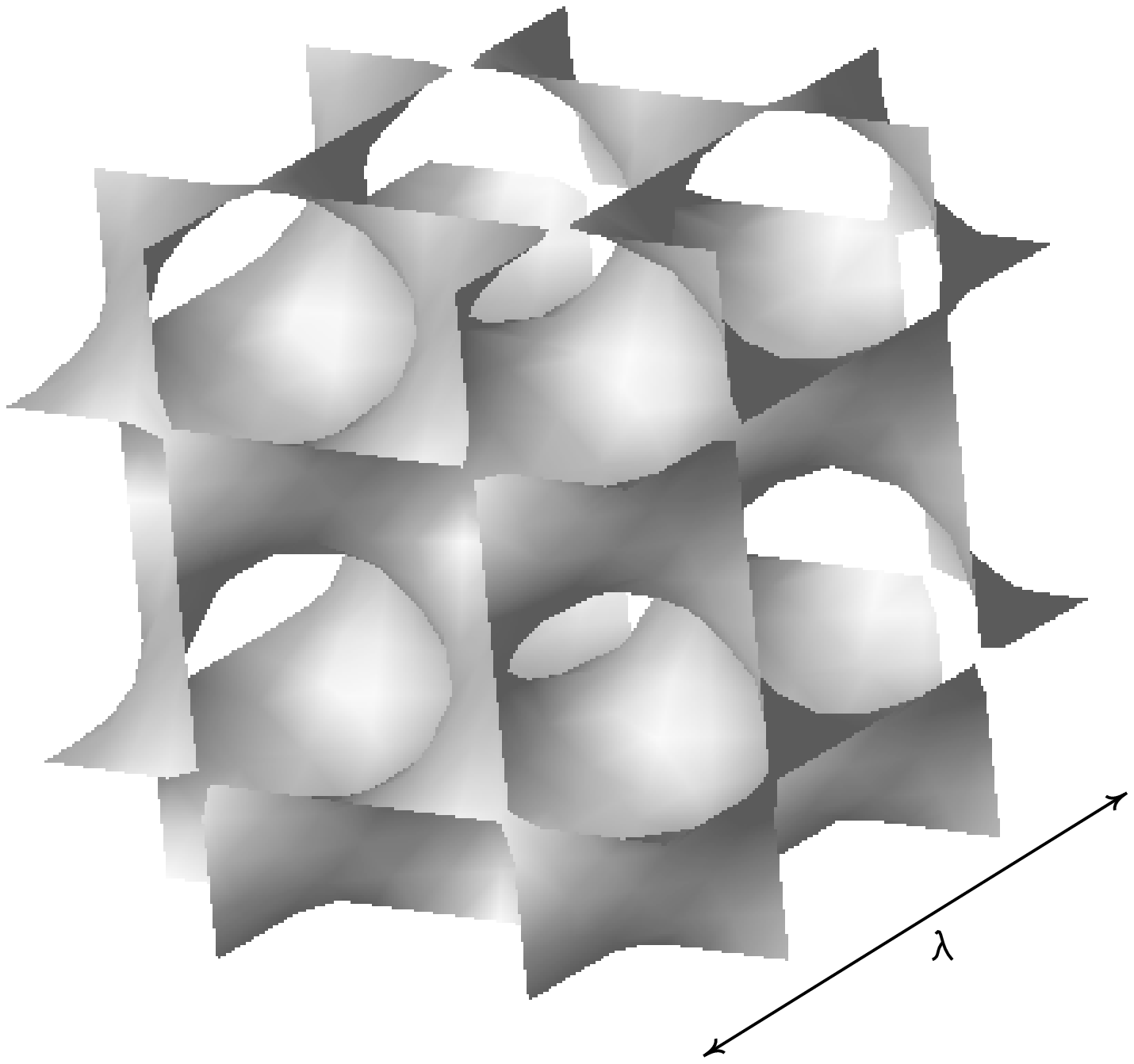}
\end{center}

\pagebreak
\LARGE
\begin{center}Fig.3\end{center}
\vskip 1cm

\begin{texdraw}
\def\water #1
{
    \bsegment
	\setsegscale{#1}
	\fcir f:0.95 r:0.8
	\larc r:0.8 sd:0 ed:360
    \esegment
}
\def\oil #1
{
    \bsegment
	\setsegscale{#1}
	\fcir f:0.2 r:0.8
	\larc r:0.8 sd:0 ed:360
    \esegment
}
\def\sxp #1
{
    \bsegment
	\setsegscale {#1}
	\arrowheadtype t:V
	\arrowheadsize l:0.5 w:0.3
	\rmove (-0.8 0)
	\ravec (1.6 0)
    \esegment
}
\def\sxm #1
{
    \bsegment
	\setsegscale {#1}
	\arrowheadtype t:V
	\arrowheadsize l:0.5 w:0.3
	\rmove (0.8 0)
	\ravec (-1.6 0)
    \esegment
}
\def\sym #1
{
    \bsegment
	\setsegscale {#1}
	\arrowheadtype t:V
	\arrowheadsize l:0.5 w:0.3
	\rmove (0 0.8)
	\ravec (0 -1.6)
    \esegment
}
\def\lattice #1
{
    \bsegment
	\setsegscale {#1}
	% drawing lattice
	\linewd 0.05
	\setgray 0.7
	\rlvec (0 8)
	\rlvec (10 0)
	\rlvec (0 -8)
	\rlvec (-10 0)
	\rmove (2 0)
	\rlvec (0 8)
	\rmove (2 0)
	\rlvec (0 -8)
	\rmove (2 0)
	\rlvec (0 8)
	\rmove (2 0)
	\rlvec (0 -8)
	\rmove (2 2)
	\rlvec (-10 0)
	\rmove (0 2)
	\rlvec (10 0)
	\rmove (0 2)
	\rlvec (-10 0)
	\rmove (0 -6)
	% filling lattice by primitives
	\setgray 0.0
	\linewd 0.1
	\rmove (1 1)\oil{#1}
	\rmove (2 0)\oil{#1}
	\rmove (2 0)\oil{#1}
	\rmove (2 0)\sym{#1}
	\rmove (2 0)\oil{#1}
	\rmove (0 2)\sxp{#1}
	\rmove (-2 0)\water{#1}
	\rmove (-2 0)\sxm{#1}
	\rmove (-2 0)\sym{#1}
	\rmove (-2 0)\sym{#1}
	\rmove (0 2)\water{#1}
	\rmove (2 0)\water{#1}
	\rmove (2 0)\sxp{#1}
	\rmove (2 0)\oil{#1}
	\rmove (2 0)\oil{#1}
	\rmove (0 2)\oil{#1}
	\rmove (-2 0)\oil{#1}
	\rmove (-2 0)\sxp{#1}
	\rmove (-2 0)\water{#1}
	\rmove (-2 0)\water{#1}
    \esegment
}
% primitives captions
\def\primcap #1
{
    \bsegment
	\setsegscale {#1}
	\linewd 0.1
	\textref h:L v:C
	\water{#1}
	\rmove (1.5 0)
	\htext {Water}
	\rmove (-1.5 -2)
	\oil{#1}
	\rmove (1.5 0)
	\htext {Oil}
	\rmove (-1.5 -2)
	\sxp{#1}
	\rmove (1.5 0)
	\htext {Oriented surfactant}
    \esegment
}
\def\toppicture #1
{
    \bsegment
	\setsegscale {#1}
	\lattice{#1}
	\rmove (14 6)
	\primcap{#1}
    \esegment
}
\def\ww #1
{
    \bsegment
	\setsegscale {#1}
	\textref h:C v:C
	\linewd 0.1
	\setgray 0.0
	\water{#1}
	\rmove (2 0)
	\water{#1}
	\rmove (-1 -2)
	\htext {$-b$}
    \esegment
}
\def\ws #1
{
    \bsegment
	\setsegscale {#1}
	\textref h:C v:C
	\linewd 0.05
	\lpatt (0.1 0.2)
	\rlvec (3 0)
	\rmove (-2.5 -1.2)
	\rlvec (2.4 2.4)
	\rmove (-1.9 -1.9)
	\linewd 0.1
	\lpatt ()
	\setgray 0.0
	\arrowheadtype t:V
	\arrowheadsize l:0.5 w:0.3
	\ravec (1.4 1.4)
	\rmove (0.6 -0.1)
	\htext {$\vartheta$}
	\rmove (-0.6 0.1)
	\rmove (-2.5 -0.7)
	\water{#1}
	\rmove (1 -2)
	\htext {$c\ \cos\vartheta$}
    \esegment
}
\def\ss #1
{
    \bsegment
	\setsegscale {#1}
	\linewd 0.05
	\rlvec (9 3)
	\rmove (-7 -2.333333333)
	\rmove (0 -2)
	\rlvec (0 4)
	\rlvec (5 1.666666666)
	\rlvec (0 -2)
	\rmove (-5 -1.66666666)
	\linewd 0.03
	\larc r:1.0 sd:270 ed:330
	\linewd 0.05
	\rmove (1.732050808 -0.866025404)
	\rlvec (-3.464101615 1.732050808)
	\rmove (3.464101615 -1.732050808)
	\rlvec (5 1.666666666)
	\rlvec (-1.732050808 0.866025404)
	\lpatt (0.1 0.2)
	\rlvec (0 -1.443375673)
	\lpatt ()
	\rlvec (0 -0.556624327)
	\rlvec (-5 -1.666666666)
	\rmove (-1.732050808 2.866025404)
	\rlvec (1.732050808 0.577350269)
	\lpatt (0.1 0.2)
	\rlvec (3.267949192 1.089316397)
	\rlvec (1.732050808 -0.866025404)
	\lpatt ()
	\rmove (-1.5 -0.5)
	\linewd 0.1
	\arrowheadtype t:V
	\arrowheadsize l:0.5 w:0.3
	\ravec (0.4 1.4)
	\rmove (-0.4 -1.4)
	\lpatt (0.03 0.15)
	\rlvec (-0.4 -1.4)
	\rmove (0.4 1.4)
	\rmove (-2 -0.666666666)
	\ravec (-1.4 0.4)
	\rmove (1.4 -0.4)
	\lpatt ()
	\rlvec (1.4 -0.4)
	\linewd 0.03
	\rmove (-1.1 -0.1)
	\larc r:1.3 sd:148 ed:173
	\larc r:1.15 sd:146 ed:171
	\rmove (1.2 0.8)
	\larc r:1.4 sd:26 ed:58
	\larc r:1.25 sd:27 ed:55
	\textref h:C v:C
	\rmove (3 2.3)
	\htext {$\vartheta_2$}
	\rmove (-7 -1.2)
	\htext {$\vartheta_1$}
	\rmove (-0.2 -3)
	\htext {$\varphi$}
	\rmove (4 -2.5)
	\htext {$-g\ \cos\varphi\sin\vartheta_1\sin\vartheta_2$}
    \esegment
}
\def\interactions #1
{
    \bsegment
	\setsegscale {#1}
	
	\ww {#1}
	\rmove (6 0)
	\ws {#1}
	\rmove (6 -1)
	\ss {#1}
    \esegment
}
\def\holepicture #1
{
    \bsegment
	\setsegscale {#1}
	\toppicture{#1}
	\textref h:L v:C
	\rmove (2 -5)
	\interactions {#1}
    \esegment
}
%
%  Start Drawing
%
\drawdim {cm}
\holepicture{0.6}
\end{texdraw}

\pagebreak
\begin{center}
Fig.4

\includegraphics[scale=1.0]{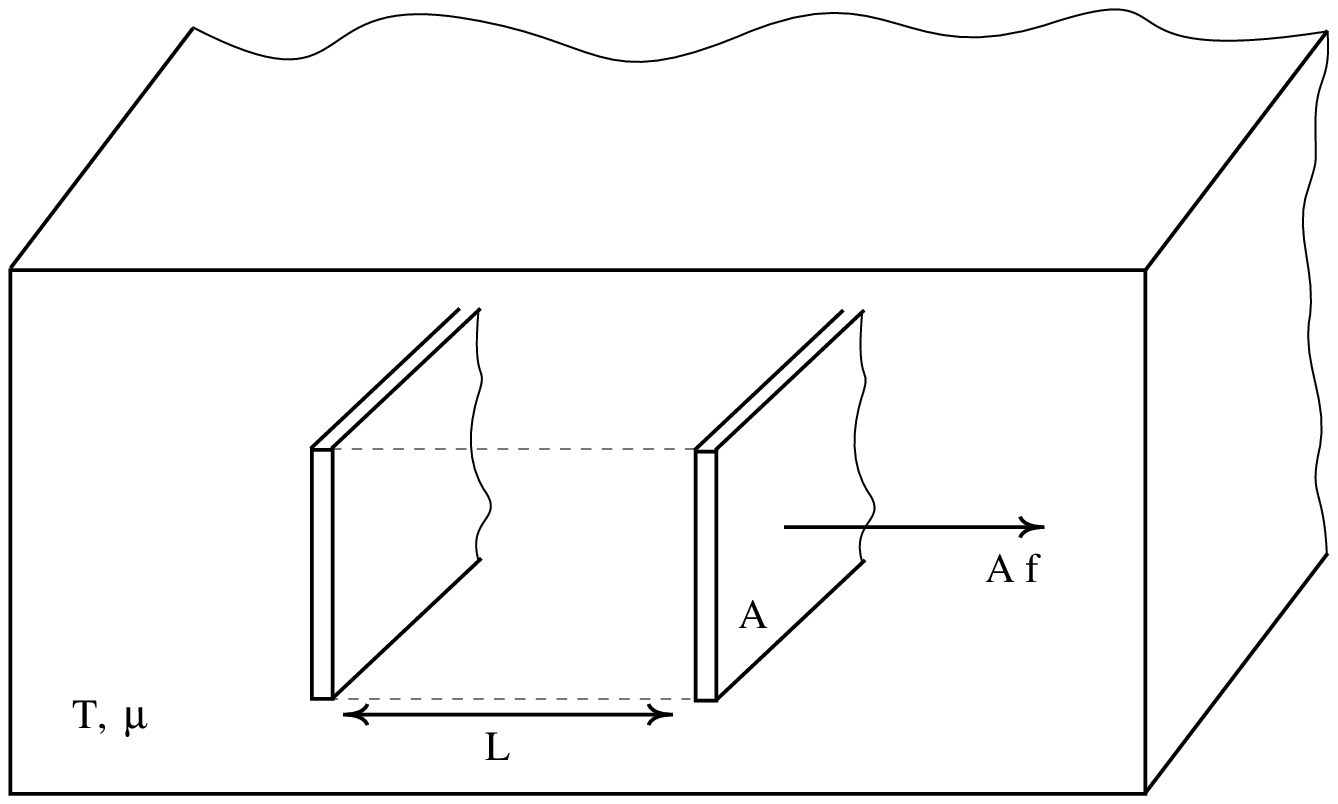}
\end{center}

\pagebreak
\begin{center}
Fig.5

\includegraphics[scale=0.7, angle=-90]{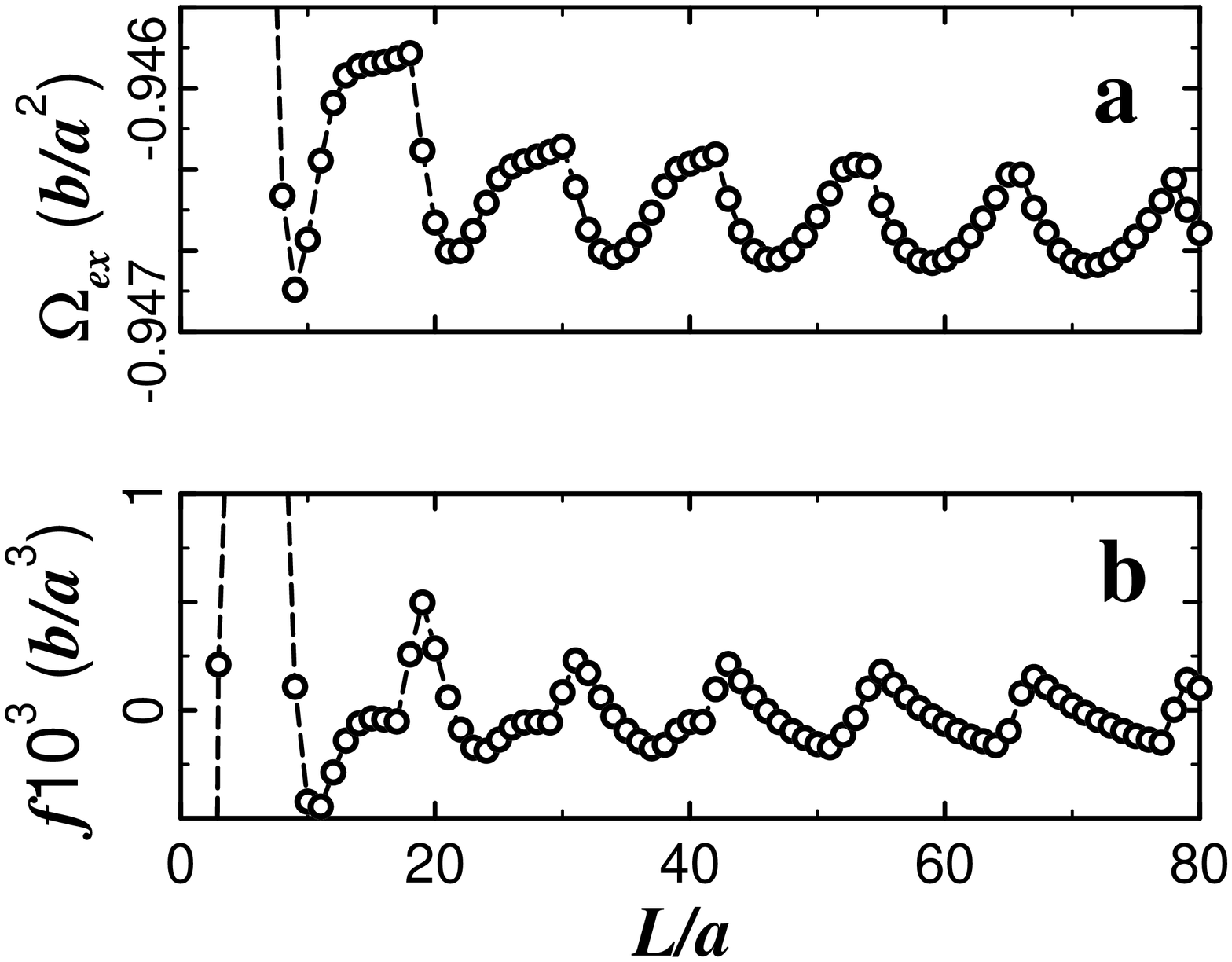}
\end{center}

\pagebreak
\begin{center}
Fig.6

\includegraphics[scale=0.7, angle=-90]{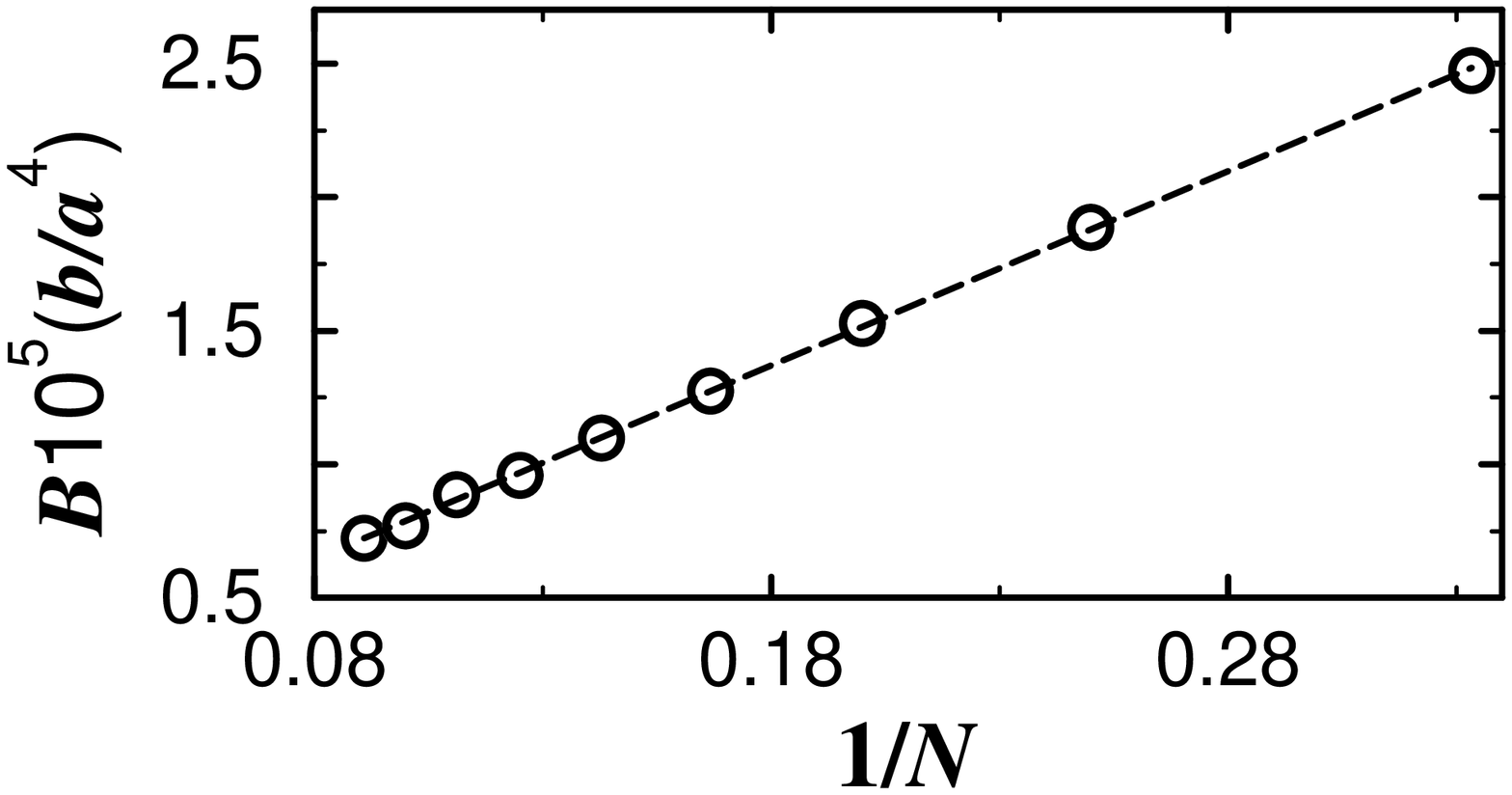}
\end{center}

\pagebreak
\begin{center}
Fig.7

\includegraphics[scale=0.7, angle=-90]{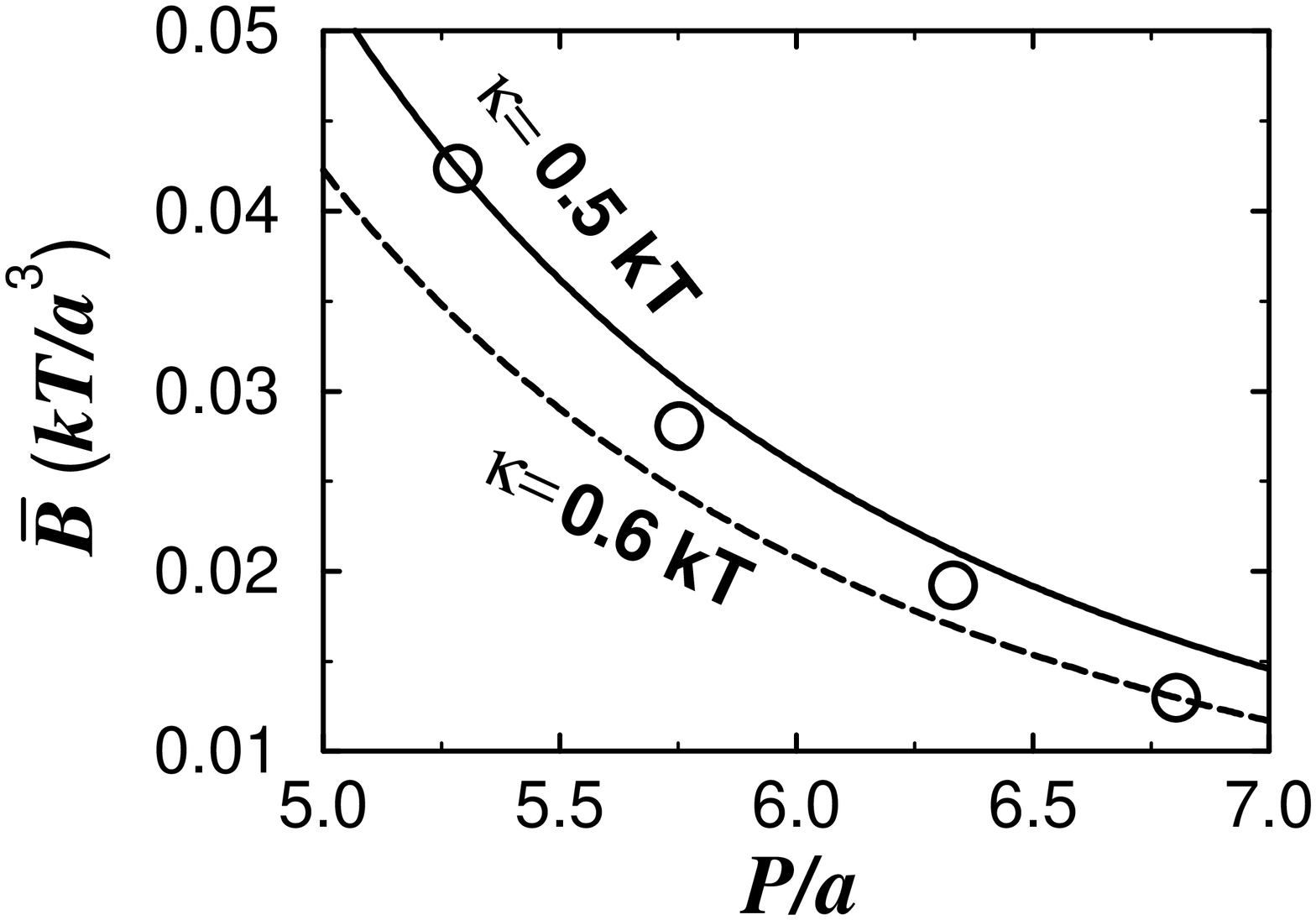}
\end{center}

\pagebreak
\begin{center}
Fig.8

\includegraphics[scale=0.7, angle=-90]{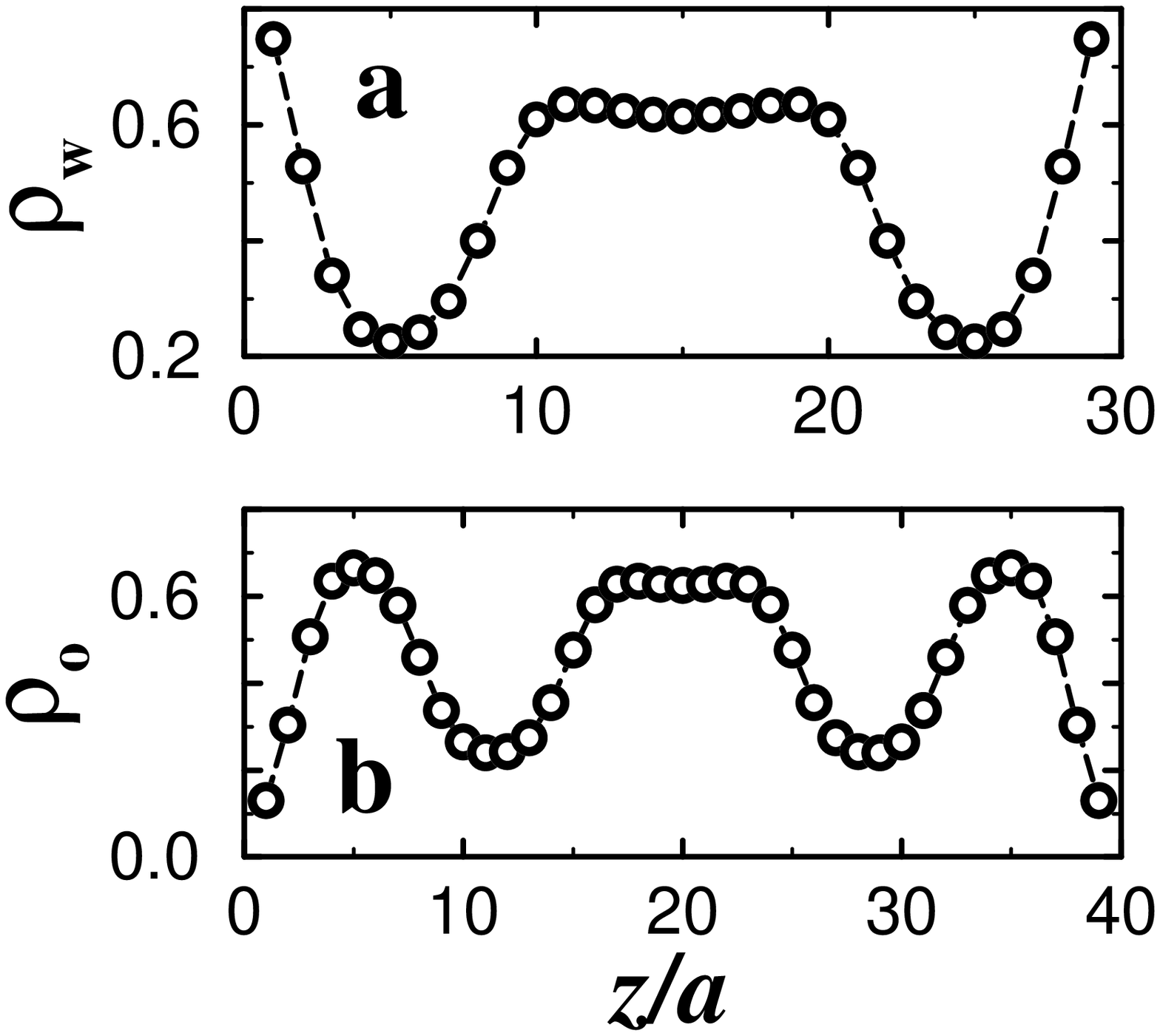}
\end{center}

\pagebreak
\begin{center}
Fig.9a

\includegraphics[scale=0.7]{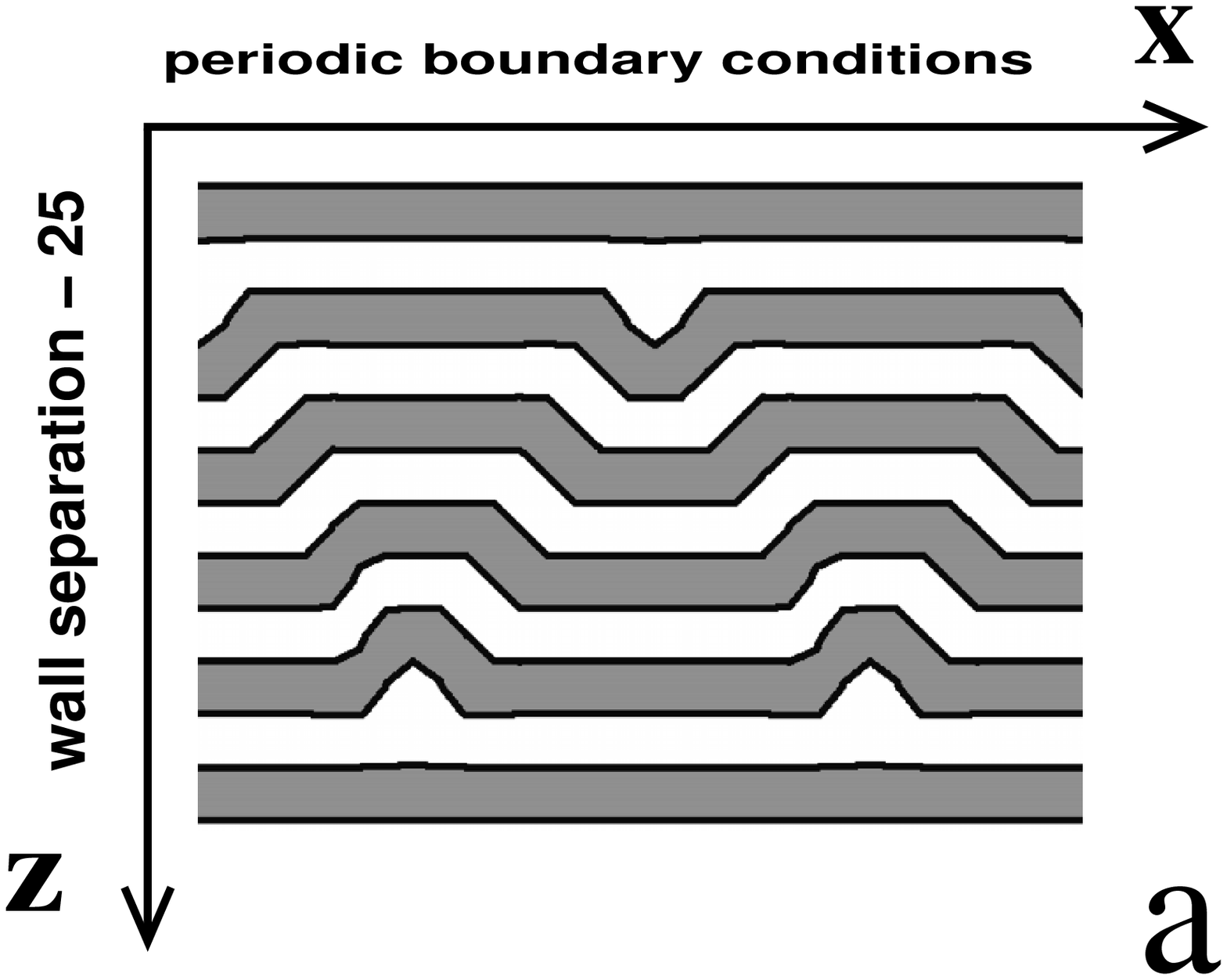}
\end{center}

\pagebreak
\begin{center}
Fig.9b

\includegraphics[scale=0.7]{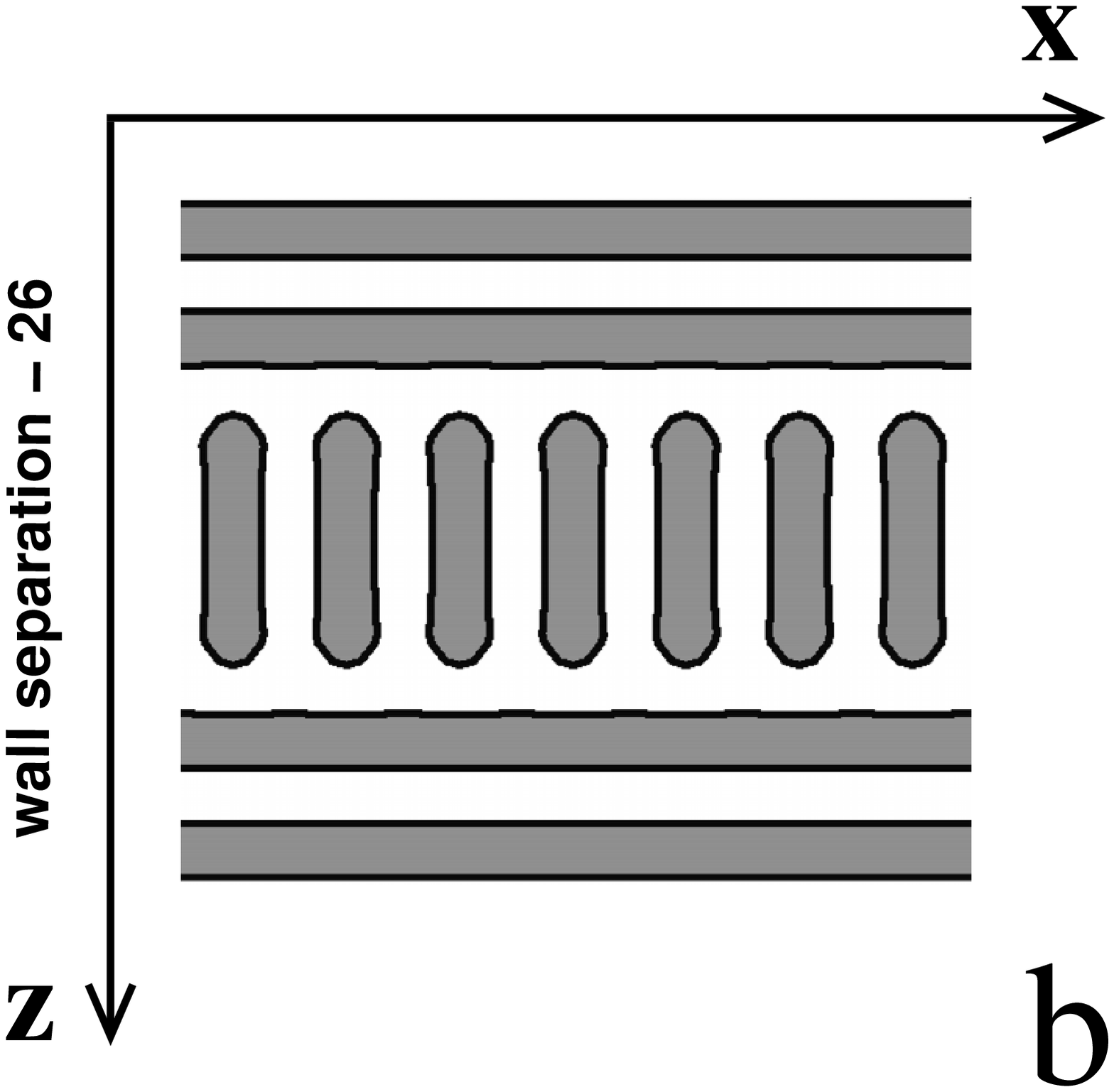}
\end{center}

\pagebreak
\begin{center}
Fig.10

\includegraphics[scale=0.8]{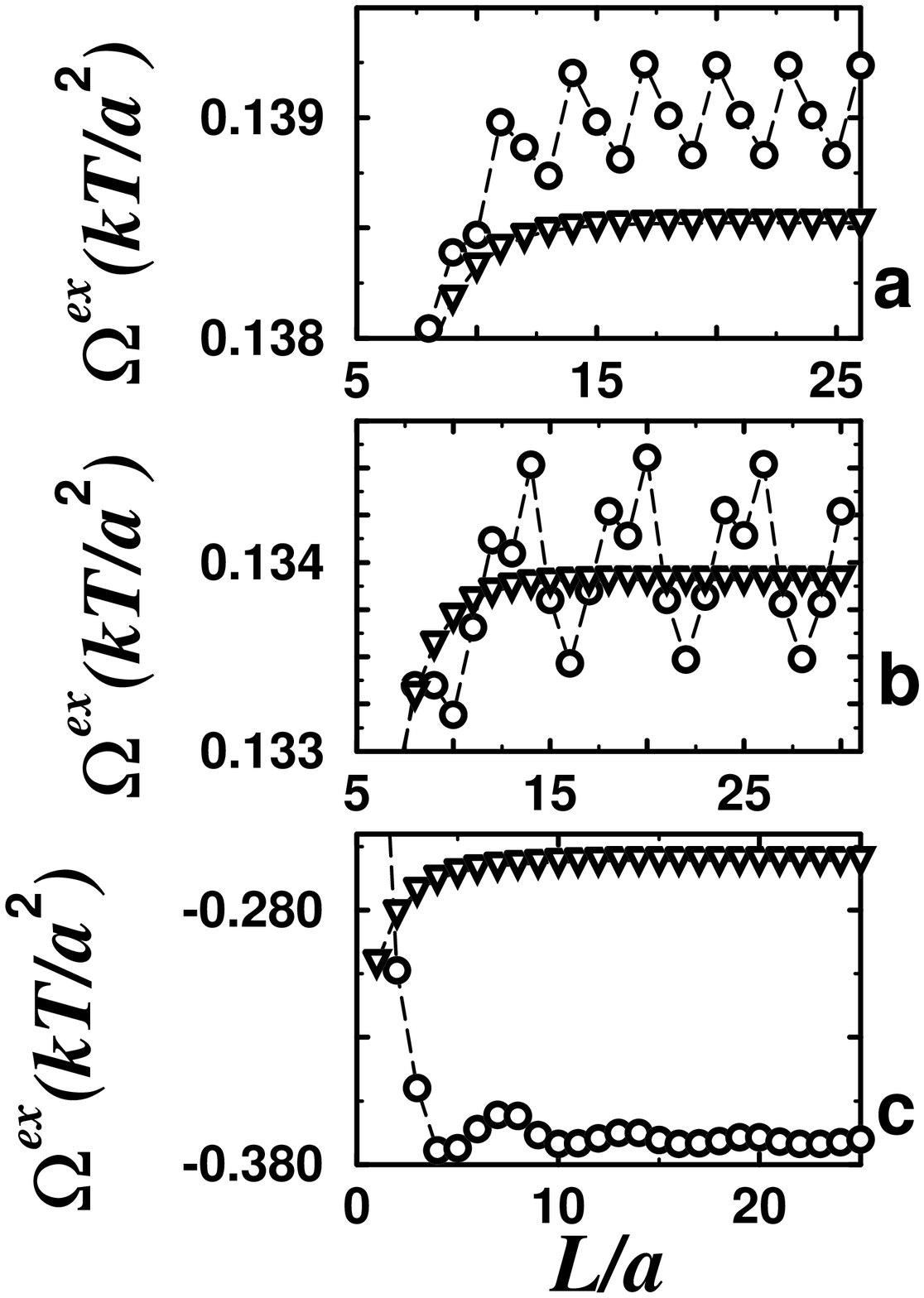}
\end{center}

\pagebreak
\begin{center}
Fig.11

\includegraphics[scale=1.1]{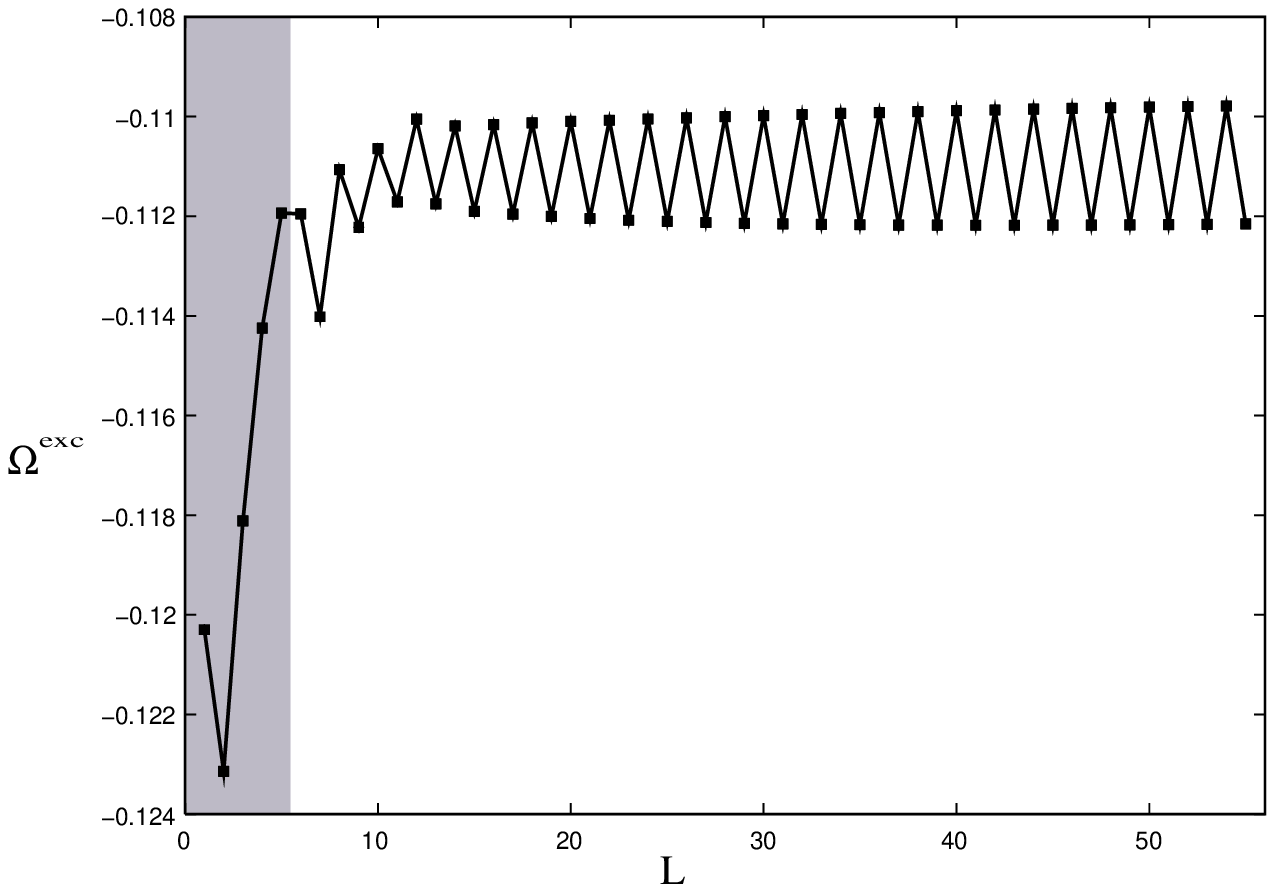}
\end{center}

\pagebreak
\begin{center}
Fig.12

\includegraphics[scale=0.9]{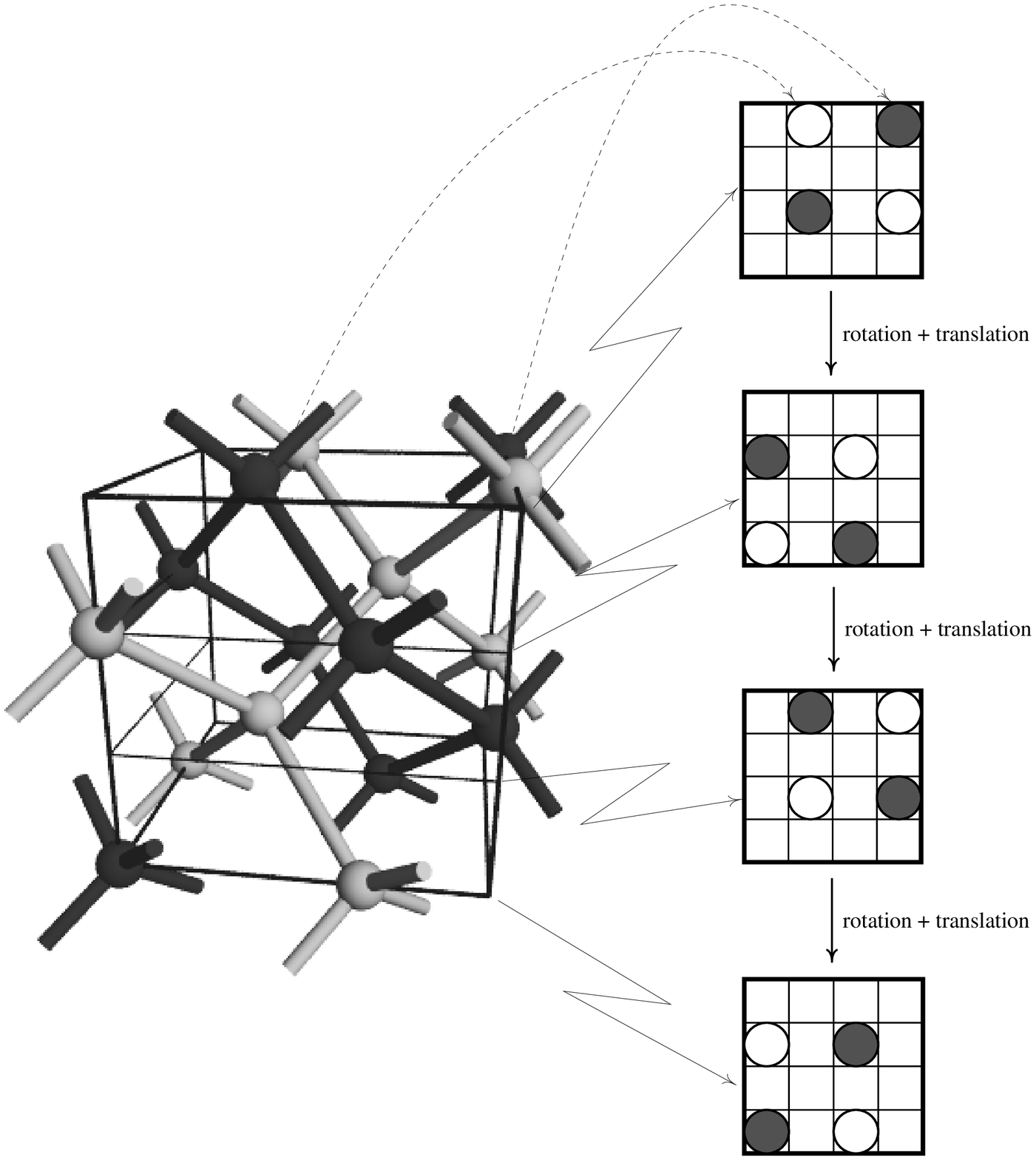}
\end{center}

\pagebreak
\begin{center}
Fig.13

\includegraphics[scale=1.0]{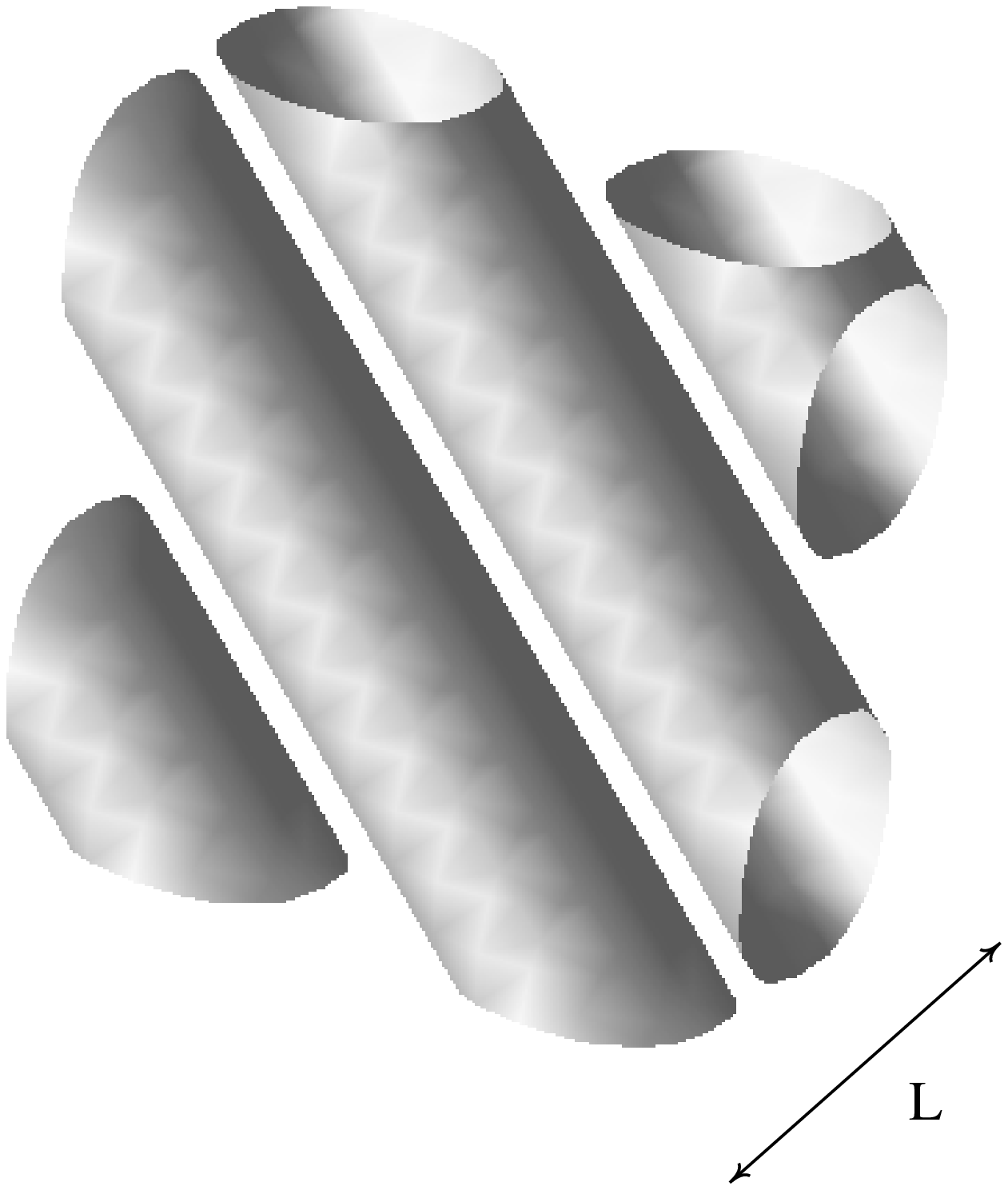}
\end{center}

\pagebreak
\begin{center}
Fig.14

\includegraphics[scale=1.0]{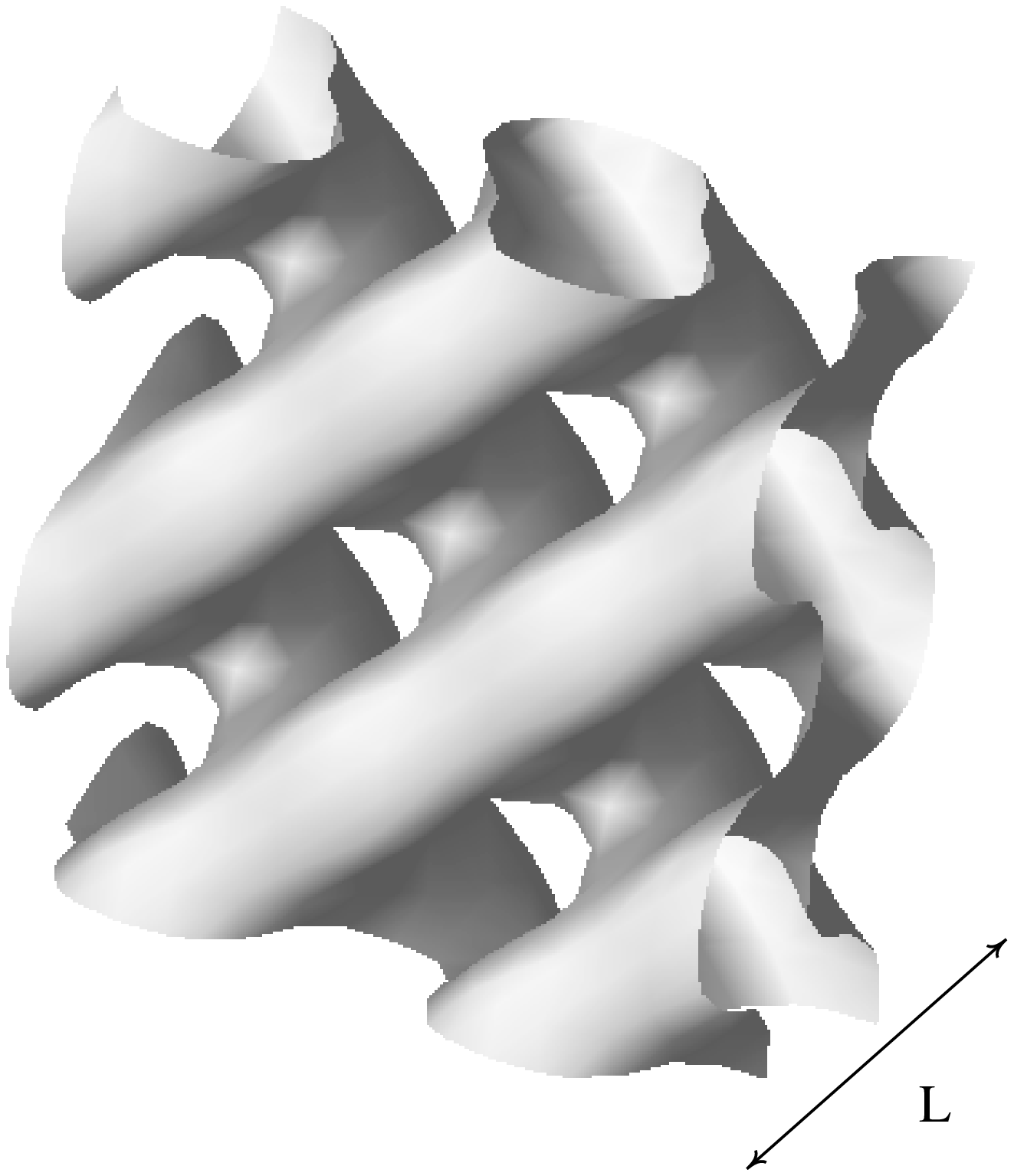}
\end{center}

\end{document}